\documentclass[a4paper,11pt]{article}

\usepackage{mathtools}
\usepackage{amssymb,bm,bbold}
\usepackage{enumerate}
\usepackage{comment}
\usepackage{esvect}
\usepackage{braket}
\usepackage[hidelinks]{hyperref}
\usepackage{a4wide}
\usepackage{cite}
\usepackage{import}
\usepackage{tikz}
\usetikzlibrary{arrows,decorations.markings}
\usepackage{tikz-3dplot}
\usepackage{mathrsfs}

\DeclareFontFamily{OT1}{pzc}{}
\DeclareFontShape{OT1}{pzc}{m}{it}{<-> s * [1.10] pzcmi7t}{}
\DeclareMathAlphabet{\mathpzc}{OT1}{pzc}{m}{it}

\newcommand{\Z}{\mathcal{Z}}
\newcommand{\cA}{\mathcal{A}}
\newcommand{\sA}{\mathscr{A}}

\newcommand{\sH}{\mathscr{H}}
\newcommand{\cH}{\mathcal{H}}

\newcommand{\sh}{\mathpzc{h}}

\newcommand{\cB}{\mathcal{B}}
\newcommand{\cN}{\mathcal{N}}
\newcommand{\cC}{\mathcal{C}}
\newcommand{\cR}{\mathcal{R}}
\newcommand{\dd}{\text{d}}

\begin{document}
	
	\numberwithin{equation}{section}
	
	\null\vskip-12pt \hfill  \\
	\null\vskip-12pt \hfill   \\
	
	\begin{flushright}
		SAGEX-21-12-E
	\end{flushright}

	\vskip2.2truecm
	\begin{center}
		\vskip 0.2truecm {\Large\bf
			{\Large Amplituhedron-like geometries}
		}\\
		\vskip 1truecm
		{\bf Gabriele Dian and Paul Heslop
		}
		
		\vskip 0.4truecm
		
		{\it
			Mathematics Department, Durham University, \\
			Science Laboratories, South Rd, Durham DH1 3LE, \vskip .2truecm                        }
	\end{center}
	
	\vskip 1truecm %

	\begin{abstract}

		We consider amplituhedron-like geometries which are defined in a similar way to the intrinsic definition of the amplituhedron but with non-maximal winding number. 
		We propose that for the cases with minimal number of points the canonical form of these geometries corresponds to the product of parity conjugate amplitudes at tree as well as loop level.  
		The product of amplitudes in superspace lifts to a star product in bosonised superspace which we give a precise definition of.
		We give an alternative definition of amplituhedron-like geometries, analogous to the original amplituhedron definition, and also a characterisation as a sum over pairs of on-shell diagrams that we use to prove the conjecture at tree level.
		The union of all amplituhedron-like geometries has a very simple definition given by only physical inequalities. Although such a union does not give a positive geometry, a natural extension of the standard definition of canonical form, the globally oriented canonical form, acts on this union and gives the square of the amplitude.

	\end{abstract}

	\medskip
	
	\noindent

	\newpage
	\tableofcontents

\section{Introduction}

In the last few years beautiful relations between scattering amplitudes and geometrical objects called positive geometries have been uncovered. The first examples of this connection~\cite{Hodges:2009hk,ArkaniHamed:2010gg} led to the discovery of the {\em amplituhedron} \cite{Arkani-Hamed2014}.
Physical observables are extracted from positive geometries by computing their canonical form, a unique top dimensional rational form with logarithmic singularities on the boundaries of the geometry~\cite{Arkani-Hamed2017}.
The canonical form of the amplituhedron is conjectured to correspond to colour ordered superamplitudes  in  planar $N=4$ SYM. A $n$ particle superamplitude is a polynomial in Grassmann odd variables where the coefficients are ordinary amplitudes, so that the information about scattering amplitudes for any particle is efficiently packaged in a single manifestly supersymmetric invariant object.  The canonical form of the amplituhedron leads to a bosonised superamplitude~\cite{Hodges:2009hk}.

In \cite{Eden2017},  a geometrical object called the {\em correlahedron} was introduced, and conjectured to be equivalent to the correlator of stress-energy tensor supermultiplets
in planar $\cN=4$ SYM. The correlahedron is not itself a positive geometry, but  evidence was given that it nevertheless possesses a well-defined volume form that should yield the correlator. This form shares fundamental properties with the canonical form, like having $\dd \log$ singularities on codimension-1  boundaries and it has been computed from the geometry for the 5 points NMHV correlator. This seems to point to some sort of generalized canonical form such as recently investigated in \cite{Benincasa:2019vqr,Benincasa:2020uph} for example.
Correlators in planar $N=4$ SYM have very direct and surprising connections with scattering amplitudes. In particular, the square of the superamplitude at all loops can be obtained as a limit of the tree-level correlator \cite{Alday_2011, Eden_2011,Adamo_2011,Eden_2012,Eden_2013,Eden_2013b}. The same limit performed geometrically on the correlahedron defines a new geometry called the squared amplituhedron, whose canonical form is thus conjectured to correspond to the square of the superamplitude. The main purpose of this work is to further investigate the squared amplituhedron conjecture and understand its geometric structure as well as investigate alternative geometries similar to the amplituhedron. Along the way we will prove at tree-level in all cases with minimal number of points, that the squared amplituhedron indeed gives the square of the amplitude, but will find that there are subtleties to  be understood in cases of non-minimal number of points.

A new definition of the amplituhedron was proposed in \cite{Arkani-Hamed2018} as a Grassmannian polytope with positive proper boundaries and maximal flipping (or winding) number, a topological invariant connected to the ordering of the kinematical data once projected down to one dimension. As pointed out there, this definition suggests a natural generalization of the amplituhedron geometry by allowing the flipping number to assume non-maximal values. We call these geometries {\em amplituhedron-like} geometries. 

We will focus almost entirely on the case of amplituhedron-like geometries with minimal number of points (maximal MHV degree) ie $k=n{-}m$ where $m=4$ in the physically interesting case but we often consider general  $m$ also. 
This is a big simplification and in particular means that the external data is trivialised. For the amplituhedron itself this case corresponds simply to the anti-MHV amplitude. However for amplituhedron-like geometries there is a very rich structure even in this sector. It contains all amplitudes multiplied by their parity conjugate amplitudes, but there is evidence that the individual amplitudes themselves can be extracted from 
this combination~\cite{Heslop:2018zut}. Furthermore this sector corresponds to taking various light-like limits of four-point correlators about which there is a wealth of concrete information. Their integrands have a hidden permutation symmetry~\cite{Eden:2011we}  and this has helped obtain their explicit expression up to ten loops~\cite{Eden:2012tu,Bourjaily:2015bpz,Bourjaily:2016evz}.

We will use the following notation to distinguish between geometrical regions, the corresponding expression in bosonised superspace, and the corresponding expression in superspace:

\begin{tabular}{c|ccc}
	&geometry & bosonised superspace & superspace   \\\hline
	amplituhedron & $\sA_{n,k,l}$ &$A_{n,k,l}$  & $\cA_{n,k,l}$\\
	amplituhedron-like &$\sH_{n,k,l}^{(f;l')}$ &$H_{n,k,l}^{(f;l')}$
	& $\cH_{n,k,l}^{(f;l')}$\end{tabular}

\vskip.3cm

\noindent    So in particular the expression in bosonised superspace is obtained from the geometry by taking the canonical form $A_{n,k,l}=\Omega(\sA_{n,k,l})$ and $H_{n,k,l}^{(f;l')}=\Omega(\sH_{n,k,l}^{(f;l')})$ and the expression in superspace is obtained from the expression in bosonised superspace by integrating out some fermionic degrees of freedom.

The squared amplituhedron of~\cite{Eden2017} is a similar geometry to the amplituhedron-like  geometry, constrained just by proper boundary inequalities but with no version of the winding condition and it can thus be viewed as the union of all amplituhedron-like geometries. 
The square of the superamplitude with fixed MHV degree $k=n-4$ is given by the sum over $k'$ of the product of the N${}^{k'}$MHV amplitude, $\cA_{n,k'}$ and its  conjugate
$\cA_{n,n-k'-4}$. 
The number of terms in this sum coincides precisely with  the number of inequivalent geometries triangulating the squared amplituhedron. It is thus natural to propose a precise relation namely that:
the amplituhedron-like geometry with flipping number $f$, $\sH_{n,k}^{(f)}$,
gives the product of the N${}^f$MHV superamplitudes and its conjugate,
\begin{align}\label{eq1}
	H^{(f)}_{n,n-4} = A_{n,f}* A_{n,n-f-4}\ .
\end{align}
We  also make a similar proposal at loop level introducing a flipping number for loops $l'$
\begin{align}\label{eq2}
	H_{n,n-4,l}^{(k',l')}= \begin{psmallmatrix}l\\l'\end{psmallmatrix}\,A_{n,k',l'} * A_{n,n-k'-4,l-l'}\ .
\end{align}
A proposal along these lines was previously made in~\cite{Arkani-Hamed2018} for the  MHV case with arbitrary number of points $H^{(0;l')}_{n,0,l}$. At first sight this is   a different sector to the case we consider. However due to factorisation of anti-MHV amplitudes this in fact  corresponds to $H^{(f;l')}_{n,f,l}$ and we will prove the relation for this case as well as at tree level. 
Here the product of amplitudes in superspace becomes a particular combination we call the star product of bosonised superamplitudes. We will give a precise definition of this star product. 

We also find an  alternative characterisation of amplituhedron-like geometries analogous to the original definition of the amplituhedron. Tree-level amplituhedron-like geometries 
with flipping number $f$ are given  in terms of a subset of the set of  matrices $C \in Gr(k,n)$ projected through $Z$. However rather than this subset of matrices $C$ having positive ordered maximal minors (which would give the amplituhedron) instead it is made up by stacking two submatrices $C_1$ (an $f\times n$ matrix) and $\text{alt}(C_2)$ (a $(k{-}f) \times n$ matrix) where $C_1$ and $C_2$ have all positive ordered maximal minors and the matrix $\text{alt}(C_2)$ is formed from $C_2$ by flipping the sign of every odd column.
A similar alternative characterisation of the amplituhedron-like geometry  can also be made at loop level.

Combinations of on-shell diagrams (arising from BCFW recursion) result in  triangulations of the amplituhedron. In a similar way we show that
at tree-level pairs of on-shell diagrams give a direct triangulation of the amplituhedron-like geometry.
This fact can then be used to prove~\eqref{eq1} for all multiplicity and winding number. We also prove the proposal at loop level in the simplest case of maximal (or equivalently minimal) flipping number $f$ giving MHV$\times$anti-MHV at specified loop levels  at all multiplicity.

Having understood the amplituhedron-like geometries it is interesting to  return to the squared amplituhedron which is the union of amplituhedron-like geometries with different flipping number.  
The square of the superamplitude shares with the superamplitude the property that it has only proper poles and dlog divergences. Differently from the superamplitude however, its maximal residues are not all normalizable to $\pm 1,0$. But this is a key property of the canonical form of any positive geometry! This therefore presents an apparent  problem in the identification of the square of the amplitude with the canonical form of the squared amplituhedron.
We solve this issue by first defining the squared amplituhedron on the {\em oriented} Grassmannian,  and then defining the {\em globally oriented canonical form}. The globally oriented canonical form coincides with the canonical form for connected geometries (and thus for the amplituhedron) but can give a different result for (almost) disconnected geometries. We then understand  the geometric origin of the non-uniform weight of the maximal residues of the superamplitude squared by the fact that is composed of the almost disconnected union of amplituhedron-like geometries, that is the union of geometries that do not share any codimension 1 boundary but do share lower codimension boundaries.
In practise, as long as the geometry is described in terms of multi-linear inequalities, the oriented canonical form can be straightforwardly evaluated using cylindrical decomposition.

Finally, all the geometries cited so far are defined by a system of inequalities depending on the kinematic data as parameters.  Thus it is interesting to see if there are any other obvious further generalisations  of the amplituhedron geometry for example by considering similar defining inequalities but with different choices of signs. As a modest step in this direction we examine carefully the consequence for such a geometry of demanding it has a manifest cyclic canonical form.  While the canonical form, i.e. the amplitude, is invariant under the rescaling of the external data $Z_i\rightarrow \lambda Z_i$, the geometry is invariant only under positive rescaling $\lambda >0$. Nevertheless, geometries related by such a transformation with $\lambda<0$ have the same canonical form. 
We thus define geometries to be {\em equivalent} if they are related by a flip of some $Z$s. 
This type of observation has already been a fundamental ingredient for proving perturbative unitarity using the amplituhedron \cite{Srikant_2020}.
Examining all possible versions of manifest geometrical cyclicity we find that all are equivalent to either cyclic or twisted cyclic geometries, thus drastically cutting down the different geometries under consideration.
As a result of this line of thinking we find an equivalence relation between amplituhedron-like geometries with complementary flipping number and new bounds for the values that they can assume. The transformation linking the two equivalent geometries corresponds to $Z_i \rightarrow (-1)^iZ_i$, a map that is closely related to parity \cite{galashin2018parity}. Using similar ideas, we consider also the maximally nilpotent correlator $\mathcal{G}_{n-4,n}$ and we prove that all the geometries with the minimal requirements to be compatible with correlator pole structure are equivalent to the correlahedron.

This paper is structured as follows. In section \ref{Thesuperamplitude} we introduce the formulation of the superamplitude in dual momentum twistor variables and we review the  bosonised superamplitude with some emphasis on its normalization. Then, we define the superamplitude squared and define a product between functions directly in the bosonised superamplitude space which we call star product and we indicate with the $*$ symbol.
In section \ref{Amplituhedron-like geometries} we define the amplituhedron-like geometries and we state our conjecture for the canonical form of amplituhedron-like geometries as products of amplitudes at tree as well as loop level. Then we define the squared amplituhedron as the union of all amplituhedron-like geometries and we conjecture that its (oriented) canonical form corresponds to the square of the superamplitude. We also give an alternative definition of the amplituhedron-like geometry as a projection of the positive and the alternating positive Grassmannian which we will then use to prove our conjecture at tree level. 
In section \ref{oriented} we review the definition of the canonical form and we introduce the oriented canonical form of a union of positive geometries as the sum of the canonical forms of the elements in the union. We then describe an algorithm called CAD to compute the oriented canonical form and which we use to perform explicit test for the oriented canonical form of amplituhedron-like geometries.
In section \ref{Proof-checks}  we show how any plane in the amplituhedron-like geometry can be seen as the product of two planes each belonging to a different amplituhedron and we use this fact along with on-shell diagrams to prove our conjecture at tree level. We then give a proof of the conjecture at all loops for the product of MHV and anti-MHV amplitudes.  We conclude the section by looking at some explicit computations for  $n\leq7$ and to some generalized amplituhedron-like geometries for $m=2,6,8$.
In section \ref{factorisation-patterns} we formulate a refined version of our conjecture for  the canonical form of regions in the amplituhedron-like geometries characterized by a precise set of inequalities called sign-flip pattern.
Finally in section \ref{CanonicalizingCyclicity} we study the equivalence relations between geometries with a cyclic canonical form and we find that for each equivalence class we can always choose cyclic or twisted cyclic representatives. We then consider the maximal nilpotent correlator $\mathcal{G}_{n,n,4}$ and prove that all consistent geometries are equivalent to the correlahedron.

\section{Amplitudes and their products in  amplituhedron space}
\label{Thesuperamplitude}

In this section we introduce the superamplitude first in dual momentum twistor superspace~\cite{Hodges:2010kq} and then its formulation in a bosonised superspace~\cite{Hodges:2009hk,Arkani-Hamed2014} which we can call ``amplituhedron space''. We note a normalisation involved in the explicit map between the two spaces. Then we focus on the meaning of the product of amplitudes directly on amplituhedron space which we denote as a $*$-product.

\subsection{Superamplitudes in momentum supertwistor space}

Planar  $n$-point  superamplitudes, ${\mathcal{A}}_n$, in $N=4$ SYM can be conveniently written as a super function of $n$ super momentum twistor variables%
\footnote{After dividing by the tree level MHV superamplitude.}
$$\mathcal{Z}_i=\begin{pmatrix} z_i \\ \chi_i \end{pmatrix} \in \mathbb{C}^{4|4}, \qquad i=1,..,n \, .$$
The momentum twistors $z_i$ are vectors in $\mathbb{C}^4$, transforming linearly under the conformal group $SU(2,2)$.
Functions of the $z_i$ invariant under the conformal group are thus naturally formed by stacking  four momentum twistors together to form a $4\times 4$ matrix and taking its determinant, denoted
\begin{align}
	\braket{z_iz_jz_kz_l}=\braket{ijkl}=\det(z_i,z_j,z_k,z_l).
	\label{zero}
\end{align}
More  generally the bracket $\braket{\cdots}$ will denote the determinant of the matrix formed by the vectors contained in the angle brackets. 
The $\chi_i$ instead are Grassmann odd variables which transform under the $SU(4)$ R-symmetry group.
The superamplitude is a singlet of $SU(4)$ which can only be obtained via contracting the $SU(4)$ indices with an $SU(4)$ $\epsilon$ tensor, thus all $\chi$ monomials in the superamplitude have a $\chi$ degree which is a multiple of four. 
Because of this, it makes sense to consider polynomials of homogeneous degree separately
\begin{align}
	{\mathcal{A}}_n= {\mathcal{A}}_{n,0}+ {\mathcal{A}}_{n,1}+ {\mathcal{A}}_{n,2}+\cdots+ {\mathcal{A}}_{n,n-4},
	\label{superamplitude}
\end{align} 
where ${\mathcal{A}}_{n,k}$ has uniform  degree $4k$ in the $\chi$'s and is called the N$^k$MHV amplitude. Here $\mathcal{A}_{n,0}=1$. 

The $\chi$ dependent building blocks have an elegant representation in terms of bosonised super-momentum twistors which  we will review now. This formulation gives a new perspective for the superamplitude that is at the core of the amplituhedron picture.

\subsection{The superamplitude in amplituhedron space}

One nice way to deal with the Grassmann odd nature of the superamplitude $\cA_{n,k}$ is to attach $4k$ additional Grassmann odd variables $\phi_{\alpha A},\ \alpha=1,..,k,\, A=1,..,4$ to each  $\chi$, thus obtain commuting variables $Z_{i \alpha}:=\chi_i^A\phi_{\alpha A}$~\cite{Hodges:2009hk,Arkani-Hamed2014} 
\begin{align}\label{zs}
	\mathcal{Z}_i= \begin{pmatrix} z_i \\ \chi_i \end{pmatrix}\quad &\rightarrow \quad Z_i(\chi_i)=\begin{pmatrix} z_i \\ \chi_i.\phi_{1} \\ \vdots \\ \chi_i.\phi_{k}, \end{pmatrix}\,
\end{align}
We then rewrite the superamplitude in terms of these bosonised supertwistors $Z_i$. More precisely we define a map $\cB_{k,4}$ from superamplitudes  (functions of $n$ momentum supertwistor space variables), to bosonised superamplitudes (functions of $n$  bosonised supertwistors in $k+4$ dimensions together with  a single $k$-plane in $k+4$ dimensions, $Y$)
\begin{align}\label{map2}
	\cB_{k,4}: \cA_{n,k}(\mathcal{Z}_i) \mapsto  A_{n,k}({Z}_i,Y)\ . 
\end{align}
The map $\cB_{k,4}$ is defined by insisting that if the bosonised $Z$s are written in terms of $\chi.\phi$ as in~\eqref{zs}, and $Y$ takes the special value $Y_0$  below,  then the result is the superamplitude times the product of all the $\phi$s:  
\begin{align}\label{ampinbos}
	A_{n,k}(Z_i(\chi_i),Y_0) &=   N(k,4) \times \prod_{\alpha=1}^k \prod_{A=1}^4 \phi_{\alpha A}\times   \cA_{n,k}(\mathcal{Z}_i) \notag \\
	Y_0&= \begin{pmatrix} 0_{4\times k} \\ \mathbb{1}_{k\times k} \end{pmatrix}\ .
\end{align}
Here $N(k,m)$ is a normalisation factor to be discussed shortly.  For now note that as long as $ A_{n,k}(Z_i(\chi),Y_0)$ is homogeneous of degree $4k$ in the $\chi$s, then it will inevitably take the form of the RHS for some function of the $\chi$s, $\cA_{n,k}(\mathcal{Z}_i)$, due to the Grassmann nature of the $\phi$s.%
\footnote{Note that the relation is more commonly written in the form $
	A_{n,k}(Z_i) =  N(k,4) \int d^{4k}\phi    A_{n,k}(\mathcal{Z})$
	which is implied by~\eqref{ampinbos} but is not as strong, since $A_{n,k}(\mathcal{Z})$ could have terms of lower degree in the $\phi$s and still satisfy this integral form. Furthermore we have not seen explicit mention of the normalisation $N(k,m)$ in the literature.
}

Since  the bosonised $\chi$s are obtained as a product of Grassmann odd quantities, they will satisfy various non-trivial nilpotency relations between them (eg $(Z_{i\alpha})^5 = 0$) which means that~\eqref{ampinbos} does not uniquely define the form of the bosonised superamplitude  $A_{n,k}(Z_i,Y_0)$ if we think of it as an ordinary function of complex variables. However the claim is that it does have a unique form with a given structure involving an emergent  $SL(4+k)$ symmetry. In particular 
a generic N${}^k$MHV-type  dual superconformal invariant can be written in a manifestly $SL(4+k)$ invariant form as the product of  $4$, $(k+4)$-brackets\footnote{In fact one can always write it as a single bracket to the power of $m$, $\braket{I_1}$, but it is useful to consider the more general case.  }

\begin{align}
	\braket{I_1}\braket{I_2}\braket{I_3} \braket{I_4},
	\label{bosInvariants}
\end{align}
where here and in the following we will use a short-hand  notation $I,J$ etc to represent an ordered set of particle numbers.
We define   $[n]:=\{1,2,...,n\}$ and then $\begin{psmallmatrix} [n] \\ k\end{psmallmatrix}$ to be the set of all ordered sets of $k$ elements in $[n]$. 
So here
$I_a\in \begin{psmallmatrix} [k{+}4] \\ k\end{psmallmatrix}$.  Any bosonized superamplitude can be written as a sum of terms of the form \eqref{bosInvariants} times a rational function of the ordinary 4-momentum twistors.

Furthermore 4-brackets involving  twistors~\eqref{zero}  can also be promoted to $(k+4)$-brackets of bosonised supertwistors by including the $(4{+}k)\times k$ matrix $Y_0$,  via the identity
\begin{align}
	\braket{ijkl}=\braket{Y_0ijkl}, \qquad \text{with}\quad Y_0= \begin{pmatrix} 0_{4\times k} \\ \mathbb{1_{k\times k}} \end{pmatrix}.
	\label{Yzero}
\end{align}
Then there appears to be a unique way of writing a function   $A_{n,k}(Z_i,Y)$ which satisfies~\eqref{ampinbos}, and which has manifest $SL(4+k)$ symmetry.

Let us consider a simple example to illustrate this.
The 5-point NMHV superamplitude  $\mathcal{A}_{5,1}$ has the form
\begin{align}
	\mathcal{A}_{5,1}(\mathcal{Z}_i)= \frac{\delta^4(\chi_1\langle2345\rangle + \text{cyclic})}{\braket{1234}\braket{2345}\braket{3451}\braket{4512}\braket{5123}}
\end{align}
and the corresponding bosonised superamplitude has the form
\begin{align}\label{nmhvinv}
	A_{5,1}(Z_i,Y_0)= [12345]:=\frac{\braket{12345}^4}{\braket{Y_01234}\braket{Y_02345}\braket{Y_03451}\braket{Y_04512}\braket{Y_05123}}\,.
\end{align}
The amplitude $A_{5,1}$ now manifests fully the SL$(k+4)$  symmetry if we allow the symmetry to  act on $Y_0$ as well as the $Z$s.
It is also straightforward to check that it satisfies~\eqref{ampinbos} with $N(1,4)=4!$. 
We  therefore  treat the $Z$s as projective vectors in $\mathbb{P}^{k+4}$, promote $Y_0$ from a constant to a variable $Y \in Gr(k,k+4)$ and study the analytic properties of $A_{n,k}(Z,Y)$.  

We will  generalize this construction, as is by now standard, by considering the momentum twistor dimension and the $\chi$s $R$-symmetry index dimension instead to be a generic positive integer $m$ rather than 4. Then a generic invariant is expressed on a $k+m$ dimensional bosonized space and will read
\begin{align}\label{inv}
	\braket{I_1}\braket{I_2}\dots \braket{I_m},
\end{align}
where $I_a\in \begin{psmallmatrix} [k{+}m] \\ k\end{psmallmatrix}$.

It is quite natural to further view this bosonised amplitude as a top form of the Grassmannian $Gr(k,k{+}m)$ that $Y$ is  an element of.
The dimension of Gr$(k,k+m)$ is $mk$, so the amplitude will be a $4k$ differential form on the Grassmannian. 
This measure has the covariant form 
\begin{align}
	\prod_{a=1}^k \braket{Y\dd^mY_{a}},
\end{align}
where $Y_a$ indicates the $a^{th}$ column of $Y$.
Notice that the measure has weight $k(m+k)$ in $Y$ and thus attaching this to the amplitude it will have weight 0 in the $Y$s as well as the $Z$s.

This construction also extends to loops (for $m=4$). In fact, for planar amplitudes there is a well defined notion of the amplitude integrand \cite{Hodges:2010kq,Mason:2010pg,ArkaniHamed:2010gh}. A loop is represented by a pair of bosonised supertwistors $(AB)$ where $A,B \in {\mathbb{C}}^{4+k}$. Bosonised amplitudes will depend on loops through the brackets $ \braket{YABZ_iZ_j}$. Its covariant measure reads $\braket{YAB\dd^2A}\braket{YAB\dd^2B}$. Loop variables always appear in the same bracket with $Y$. Therefore they are naturally defined on $Y^\perp$, and are elements of Gr$(2,4)$.

Summarising, the bosonised superamplitude $A_{n,k,l}$ can be written as a rational differential form depending on $Y\in Gr(k,k+4)$  and $l$ loop variables which are lines in $Y^\perp$ so effectively $(AB)_i\in Gr(2,4)$, together with $n$  $Z$'s in Gr$(1,k+4)$. 
Remarkably the resulting differential form is the unique canonical form obtained from a simple geometrical object, the amplituhedron. In the next section we will give a brief review of this geometrical formalism.

Finally we discuss the normalization $N(k,m)$ appearing in the map from superspace to amplituhedron space~\eqref{ampinbos}. This is present simply due to the combinatorics involved in  extracting the $\phi$s from the amplituhedron-type expression.
It can be motivated and derived through the example of the anti-MHV $k=n-4$ amplitude. This has a simple expression in amplituhedron space:
\begin{align}\label{13}
	A_{n,n-4}(Z_i,Y)= \frac{\braket{1\cdots n}^4}{\prod_i\braket{Yi(i+1)(i+2)(i+3)}}.
\end{align}
But in order for this to give the corresponding superspace expression we need to pull out the $\phi$s, yielding a numerical factor.
Explicitly then, for general $m$, the numerical factor $N(k,m)$ is fixed by
\begin{align}
	\det(\phi_{i}.\chi_j)^m=N(k,m)\prod_{i=1}^{k}\prod_{A=1}^m \phi_{i A} \prod_{i=1}^{k}\prod_{A=1}^m \chi_i^A\ .
\end{align}
So for $m=1$, for example,  we don't have any R-symmetry index and every term in the (single) determinant contributes the same giving a factor of $k!$. Taking into account the re-ordering of the Grassmann variables then gives  $N(k,1)=(-1)^{\lfloor\frac{k}{2}\rfloor}k!$. More generally, by computing a number of cases explicitly we find  they are always consistent with the following expression
\begin{align}
	N(k,m)= (-1)^{\lfloor\frac{m k}{2}\rfloor}(m!)^k \prod_{j=1}^k \frac{(m+j)^{k-j}}{j^{k-j}}.
\end{align}

\subsection{The squared superamplitude}

\label{introsqamplituhedron}

The superamplitude is a polynomial in the Grassmannian variables $\chi$ that can be organized as a sum of polynomials of uniform degree, the NMHV sectors~\eqref{superamplitude}. The same thing can be done for the product of the full superamplitude (sum over all N${}^k$MHV sectors)  with itself, the superamplitude squared, which will be the main object we will be interested in this work. Explicitly at loop level the superamplitude is
\begin{align}
	\mathcal{A}_n= \sum_{k=0}^{n-4}  \mathcal{A}_{n,k,l},
\end{align}
and so squaring this the
superamplitude squared simply reads 
\begin{align}
	\mathcal{A}_n^2= \sum_{k=0}^{n-4}\sum_{k'=0}^k \mathcal{A}_{n,k'}  \mathcal{A}_{n,k-k'},
\end{align}
where we would like to stress the fact that these products are between functions of anti-commuting variables. Each product $\mathcal{A}_{n,k'}  \mathcal{A}_{n,k-k'}$  has uniform degree in $\chi$ equal to $4(k'+(k-k'))=4k$. Therefore, we can define the superamplitude squared NMHV sectors as
\begin{align}
	(\mathcal{A}^2)_{n,k}= \sum_{k'=0}^k \mathcal{A}_{n,k'}  \mathcal{A}_{n,k-k'}.
	\label{sas}
\end{align}
Notice that each term in the sum is a product of two dual superconformal invariants and therefore $(\mathcal{A}^2)$ will also be dual superconformally invariant. This means that we can follow the same bosonisation procedure we used for the the amplitude and write $(\mathcal{A}^2)_{n,k}$ in terms of the  $k+4$ brackets \eqref{bosInvariants}. 

At loop level the amplitude is a double sum over MHV degree and loop level:
\begin{align}
	\mathcal{A}_{n}=\sum_{l=0}^\infty \sum_{k=0}^{n-4} \int d\mu_{l}\,\mathcal{A}_{n,k,l}\,.
	\label{loops}
\end{align}
Here  the amplitude $\mathcal{A}_{n,k,l}$ is symmetric respect to the loop variables $\{AB_1,\cdots,AB_l\}$ and we define the integration measure as weighted by $1/l!$ compared to $l$ copies of the  1-loop measure: 
\begin{align}
	d\mu_l[(AB)_1, .., (AB)_l]:=\frac{d\mu_1[(AB)_1] .. d\mu_1[(AB)_l]}{l!}\,.
\end{align}
Then when we take the square of the amplitude we obtain
\begin{align}
	\mathcal{A}_n^2=\sum_{l=0}^\infty\sum_{k=0}^{n-k}  \sum_{l'=0}^l \sum_{k'=0}^k  \int d\mu_{l'} d\mu_{l-l'}\mathcal{A}_{n,k',l'} \mathcal{A}_{n,k-k',l-l'}:=\sum_{l=0}^\infty\sum_{k=0}^{n-k} \int d\mu_l\,(\mathcal{A}^2)_{n,k,l} \,,
	\label{loopsa}
\end{align}
Thus by the N${}^k$MHV $l$-loop squared amplitude we mean 
\begin{align}
	(\mathcal{A}^2)_{n,k,l}=\sum_{l'=0,k'=0}^{l,k}\begin{pmatrix} l \\ l' \end{pmatrix}   \mathcal{A}_{n,k',l'} \mathcal{A}_{n,k-k',l-l'}\,.
	\label{loopsa2}
\end{align}
The numerical factor arises from matching the measures $\int d\mu_{l'} d\mu_{l-l'}= \begin{psmallmatrix} l \\ l! \end{psmallmatrix} \int d\mu_{l}$. Note however that we have not specified the distribution of the $l$ loop variables between the two factors $\mathcal{A}_{n,k',l'}$ and $\mathcal{A}_{n,k-k',l-l'}$. The most natural choice is to have a completely symmetric distribution in which case 
there are exactly $\begin{psmallmatrix} l \\ l! \end{psmallmatrix}$ inequivalent ways to do this and the squared amplitude simply sums over all these inequivalent distributions~\eqref{loopsa2}.

\subsection{The star product}

We now wish to consider the squared amplitude  in amplituhedron space. We must therefore understand the outcome of taking the product of amplitudes in amplituhedron space. Note that this can not be given simply by the product of amplitudes in amplituhedron space, as these will live in different spaces. 
Instead we  define a map we call  $*$ which takes  two  amplitudes in amplituhedron space and produces a third  amplitude in amplituhedron space which will be equivalent to  the product of the two original superamplitudes
under the map~\eqref{map2},\eqref{ampinbos}:
\begin{align}\label{diag}
	\cB_{k_1+k_2,4}\Big( \cA_{n,k_1}(\Z_i) \cA_{n,k_2}(\Z_i) \Big) =
	A_{n,k_1}(Z_i,Y_{k_1}) * A_{n,k_2}(Z_i,Y_{k_2})
	\ .
\end{align}
Note that the $*$ takes an object in  $k_1{+}4$ dimensions  and an object in $k_2{+}4$ dimensions and outputs an object in  $k_1{+}k_2{+}4$ dimensions.

We now give an explicit definition of this $*$ product via its action on arbitrary dual superconformal invariants~\eqref{inv}.
So we consider the product of two dual superconformal building blocks~\eqref{bosInvariants} of degree $k_1$ and $k_2$ respectively. 
In superspace the product is clear, but what happens in the bosonised amplituhedron space when we take the product?
Generalising to arbitrary $m$, the bosonised invariants live in  dimensions, $k_1+m$ and $k_2+m$ dimensions respectively, and  we want to write the product as an object  $k_1+k_2+m$ dimensions.
To keep track of the $\phi$ dependence we will add the subscript $k_1+m$ to the $k_1+m$ dimensional brackets and the subscript $k_2+m$ to the $k_2+m$ dimensional brackets.
We  label these brackets  by the strings $I_a \in \begin{psmallmatrix} [n] \\ k_1+m \end{psmallmatrix} $ and $J_b \in \begin{psmallmatrix} [n] \\ k_2+m \end{psmallmatrix} $.

We claim that the $*$ product of bosonized brackets is given by the formula
\begin{align}
	\left(\prod_{a=1}^m\braket{I_a}_{k_1+m}\right)*\left(\prod_{b=1}^m\braket{J_b}_{k_2+m}\right)= \frac{(-1)^{(k_1k_2+k_2) m}}{m!} \sum_{\sigma\in S_m} \prod_{a=1}^m \braket{Y (I_a\cap J_{\sigma(a)})}_{k_1+k_2+m}
	,
	\label{prodRule}
\end{align}
where $Y$ is in $Gr(k_1{+}k_2,k_1{+}k_2{+}m)$.   Here  $S_m$ is the set of permutations of $m$ elements and $(I\cap J)$ represents an intersection in $k_1+k_2+m$ dimensions, explicitly:
\begin{align}\label{intersection}
	\braket{Y (I\cap J)}=\sum_{i\in M(I)}\braket{Yi}\braket{\overline{i} J} \text{sgn}(i  \overline{i}),
\end{align}
where $M(I)= \begin{psmallmatrix} I \\ m \end{psmallmatrix} $, that is the set of ordered $m$ tuples in $I$, and $\overline{i}$ is the ordered complement of $i$ in $I$, that is $\overline{i}= I-i$.

Note that if we set $Y$ to  $Y_0$ and the $Z$s to $Z(\chi)$, as defined in \eqref{zs}, and include the normalisation factor $N(k,m)$ then the star product formula \eqref{prodRule} must  reduce to an ordinary product. (This is just from the defining equation~\eqref{diag} and the definition of the map  $\cB$~\eqref{map2},\eqref{ampinbos}).
Thus to prove the explicit form of the star product~\eqref{prodRule} we need to check that when $Z \rightarrow Z(\chi)$:
\begin{align}
	\frac{1}{N(k_1,m)}\left(\prod_{a=1}^m\braket{I_a}_{k_1+m}\right)\frac{1}{N(k_2,m)}\left(\prod_{b=1}^m\braket{J_b}_{k_2+m}\right)=\notag \\
	=\frac{1}{N(k_1+k_2,m)} \frac{(-1)^{(k_1k_2+k_2) m}}{m!} \sum_{\sigma\in S_m} \prod_{a=1}^m \braket{Y_0 (I_a\cap J_{\sigma(a)})}_{k_1+k_2+m}
	\,.
	\label{prodRule2}
\end{align}
We include the proof of this for $m=1$ and some checks for $m=2$ and $m=4$ in the appendix.

\subsubsection*{Example}

As an example, let's look at the squared amplitude $(\mathcal{A}^2)_{6,2}$. This is given by two terms
\begin{align}
	(\mathcal{A}^2)_{6,2}= 2 \mathcal{A}_{6,2}+ (\mathcal{A}_{6,1})^2\ .
	\label{A262}
\end{align}
Now we want  to express \eqref{A262} as a function on the bosonised amplituhedron superspace.
The first term is the $6$ points anti-MHV amplitude $A_{n,n-4}$ given in~\eqref{13}.
For the second term we start with the  BCFW expression for $A_{6,1}$~\cite{Drummond:2008vq} in terms of the 5-point NMHV-invariant~\eqref{nmhvinv}, that is
\begin{align}
	A_{6,1}=[12345]+[12356]+[13456].
\end{align}
To compute the square of $A_{6,1}$ we need the (star) product of  5-brackets.  Identifying a 5-bracket  as $\braket{\hat i}$, where $\hat i$ indicates the unique twistor that is not present and a 4-bracket as $\braket{\hat i \hat j}$ similarly, the star product  formula~\eqref{prodRule} gives
\begin{align}
	\braket{\hat i}^4*\braket{\hat j}^4= \braket{Y\hat i \hat j}^4 \braket{123456}^4\ .
	\label{invariantproduct}
\end{align}
Indeed, as pointed out in \cite{Heslop:2018zut}, the result is completely fixed up to proportionality  by matching the scaling in each $Z$.
The square of any $R$-invariant will be equal to zero.
We possess now all the elements to compute $(A_{6,1})^2$ and obtain
\begin{align}
	(A_{6,1})^{*2}&=2\left([12345]*[12356] +[12345]*[13456]+[13456]*[12356]\right)= \nonumber \\
	&=2 \braket{123456}^4\frac{ \braket{1245}\braket{2361}\braket{3456}+
		\braket{2356}\braket{3412}\braket{4561}+
		\braket{3461}\braket{4523}\braket{5612}}{\prod_{i=1}^3\braket{i(i{+}1)(i{+}3)(i{+}4)})\prod_{i=1}^6\braket{i(i{+}1)(i{+}2)(i{+}3)}}\,,
	\label{A61}
\end{align}
where the 4-brackets $\braket{*}$ are short-hand for $\braket{Y*}$. Summing this result with $A_{6,2}$~\eqref{13} we obtain $(A^2)_{6,2}$ in amplituhedron space.

\subsubsection*{Product of multiple amplitudes}

The product of multiple bosonised brackets can be computed just by using the associative property of the * product. However it's also possible to write a direct formula for the * product of multiple brackets. To do this notice that
\begin{align}
	\braket{Y (I\cap J)}= \braket{I (Y\cap J)}.
\end{align}
which can be checked by expanding the respective intersections on each side  out over the $J$ basis
\begin{align}
	\braket{Y(I\cap J)}= \sum_j \braket{Yj}\braket{I\bar{j}}\text{sgn}(j\cup\bar{j}) \notag \\
	\braket{I(Y\cap J)}=\sum_j \braket{I\bar{j}}\braket{Yj}\text{sgn}(\bar{j}\cup j).
\end{align}
Using this alternative expression, equation \eqref{prodRule} for the product of 2 terms naturally generalizes to the product of $t$ terms as
\begin{align}\label{multiprodRule}
	\left(\prod_{a=1}^m\braket{I_{1,a}}_{k_1+m}\right)*\cdots*\left(\prod_{a=1}^m\braket{I_{t,a}}_{k_t+m}\right)=\notag \\ = \frac{1}{(m!)^t}   \sum_{\sigma_* \in (S_m)} \prod_{a=1}^m \braket{ I_{1,a}(Y\cap I_{2,\sigma_2(a)})\cdots (Y\cap I_{t,\sigma_t(a)}) }_{k_1+\cdots+k_t+m},
\end{align}
up to a sign which is positive for $m$ even and depends on $k_1,\cdots, k_t$ for $m$ odd.

\subsection{Maximal residues of the squared amplituhedron}
\label{maxres}
Because the square of the superamplitude can be written as a sum of products of on-shell diagrams (see section \ref{On-shell diagrams})  it only has dlog singularities just like the superamplitude itself. Differently from the superamplitude however, the maximal residues of the square of the superamplitude are not all 
$\pm 1$. A simple consequence of the standard recursive definition of  the
canonical form of a positive geometry given in~\cite{Arkani-Hamed2017} is that all its non-trivial maximal residues are $\pm 1$. This looks like an apparent  problem for a geometric description of the square of the amplituhedron.
Nevertheless we will define  in section~\ref{oriented},   an alternative and quite natural generalisation of the canonical form which  we call the {\em oriented canonical form}. It can be defined on any subspace of an oriented space defined by linear inequalities. Maximal residues of this oriented canonical form can have different absolute values and we will see that it  allows for a geometrical avatar to the square of the superamplitude.

To illustrate the point about maximal residues, we give here an explicit example of two residues that have different absolute value, consider again the $n=6$,  $k=2$ superamplitude squared, that is given by \eqref{A262} lifted to amplituhedron space
\begin{align}
	({A}^2)_{6,2}= 2 {A}_{6,2}+ {A}_{6,1}* {A}_{6,1}\ .
	\label{A2622}
\end{align}
Note that a factor of $2$ is manifest in the first term but is also present in the expression for the second term \eqref{A61}. These two terms then have uniform maximal residues equal to $\pm2$ or 0. We can examine this explicitly,  using the coordinates 
\begin{align}
	Y=\left(
	\begin{array}{cccccc}
		1 & \alpha _2+\alpha _4+\alpha _6+\alpha _8 & \left(\alpha _2+\alpha _4+\alpha _6\right) \alpha _7 & \left(\alpha
		_2+\alpha _4\right) \alpha _5 & \alpha _2 \alpha _3 & 0 \\
		0 & 1 & \alpha _7 & \alpha _5 & \alpha _3 & \alpha _1 \\
	\end{array}
	\right)
\end{align}
and setting $Z=\mathbb{1}$.  These two terms then read
\begin{align}
	2A_{6,2}&=2\prod_{i=1}^8 \frac {d\alpha_i} {\alpha_i}, \\
	(A_{6,1})^2&=2 \prod_{i=1}^8 \frac {d\alpha_i} {\alpha_i}\left(1-\frac{\alpha _2 \alpha _6+\alpha_4\alpha_8}{(\alpha _4{+} \alpha _6{+} \alpha _8)(\alpha
		_2{+}\alpha _4{+}\alpha _6)} \right).
\end{align}
From this parametrized form we can see that for example both terms contribute equally to the multi-residue corresponding to sending  $\alpha_2,\alpha_4\to 0$ (in either order) and we thus have
\begin{align}
	\text{Res}_{\alpha_2,\alpha_4\to 0}(A^2)_{6,2}=4 \frac{d \alpha_1 d\alpha_3d\alpha_5d \alpha_6d \alpha_7d \alpha_8}{\alpha_1 \alpha_3\alpha _5 \alpha _6 \alpha _7 \alpha _8},
\end{align}
and will thus yield a maximal residue of 4. On the other hand, the residue corresponding  to 
first taking $\alpha_8\to 0,\alpha_4\to 0$ and then taking $\alpha_6 \to 0$  vanishes for $(A_{6,1})^2$  and so 
\begin{align}
	\text{Res}_{\alpha_8,\alpha_4,\alpha_6\to 0}(A^2)_{6,2}=2 \frac{d \alpha_1 d \alpha_2 d \alpha_3 d \alpha_5d \alpha_7}{\alpha_1\alpha_2\alpha_3 \alpha_5 \alpha_7  }
\end{align}
yielding a maximal residue of 2.
In general we will have that the maximal residues of $(A^2)_{6,2}$ are all equal to $0,\pm2$ or $\pm4$. Therefore $(A^2)_{6,2}$ can not  be interpreted as the canonical form of a positive geometry with the standard definition. But we will find it does have a very natural interpretation as the {\em oriented canonical form} of a geometry (section~\ref{oriented}).

\section{ Amplituhedron-like geometries}
\label{Amplituhedron-like geometries}

Having discussed the form of amplitudes, their products and the squared amplitude in amplituhedron space we now turn to the corresponding geometries. We first review the amplituhedron geometry~\cite{Arkani-Hamed2014,Arkani-Hamed2018} before defining a natural generalisation of this which we call ``amplituhedron-like'' geometries which we argue corresponds to the product of amplitudes, at least in the maximal $k=n-4$ case.

\subsection{The amplituhedron}

The amplituhedron is a geometrical object introduced in \cite{Arkani-Hamed2014} which is equivalent to the amplitude. Its codimension one boundaries correspond to the locus of the order one poles of the bosonised amplitude. This generalizes to lower order boundaries, so the loci of the order $p$ poles 
correspond to the codimension $p$-boundaries of the amplituhedron. 
The amplituhedron is described  $\sA_{n,k}$ is the subspace of $Gr(k,k+4)$ defined as
\begin{align}
	\text{tree amplituhedron: }\quad  \sA_{n,k}(Y;Z):= \{Y=C\cdot Z \in Gr(k,k+4) | \ C\in Gr_{>}(k,n)\}, \notag \\\quad \text{for }Z\in Gr_{>}(k+4,n)\,,
	\label{amplituhedron}
\end{align}
where $Gr_{>}(k,n)$ is the space of oriented $k$-planes for which all the maximal ordered minors are positive and is called the positive Grassmannian \cite{Arkani-Hamed:2016byb}. The positive Grassmannian is inherently real and therefore $\sA_{n,k}$ is defined as a region in the real {\em oriented} Grassmannian $\widetilde{Gr}(k,k+4):=\mathbb{R}^{k\times 4}/GL^+(k)$, that is the space of oriented $k$-planes  in $k+4$ dimensions.
The amplituhedron is usually then viewed as being the projection of this onto the (unoriented) real Grassmannian $Gr(k,k+4)$.  However we instead find it useful to remain on  $\widetilde{Gr}(k,k+4)$ and view the amplituhedron directly on this space. This allows for a natural universal orientation for any subset.
The amplitude itself is extracted from the geometry by taking its canonical form (see section~\ref{oriented}) and will therefore also initially be defined on the real Grassmannian,   but can be then analytically continued to the complex numbers. 
This definition of $Y$ through the matrix $C$ is in general degenerate, that is two different $C$'s in Gr$(k,n)$ can correspond to the same $Y\in Gr(k,k+4)$. We can write $Y$ using the $C$ matrix Pl\"ucker coordinates as
\begin{align}
	Y= \sum_{1\leq  i_1<\cdots<i_k\leq n} \det (C_{i_1},\cdots,C_{i_k})Z_{i_1}\cdots Z_{i_k},
	\label{convexity}
\end{align}
where $C_{i}$ is the ${i}$-th column of the matrix $C$. Using \eqref{convexity} we can see that the brackets $\braket{Yii+1jj+1}$ are always positive,
\begin{align}
	\braket{Yii+ijj+1}= \sum_{1\leq  i_1<\cdots<i_k\leq n} \det (C_{i_1},\cdots,C_{i_k})\braket{Z_{i_1}\cdots Z_{i_k}Z_iZ_{i+1}Z_jZ_{j+1}}>0,
\end{align}
where  we used that $Z\in Gr_{>}(k+4,n)$~\eqref{amplituhedron}. Each term in the sum is positive since it is given by the product of an ordered $C$ minor and an ordered $Z$ minor. The $j=n$ case is special and one can check that the bracket $\braket{ii+1n1}$ is positive for $k$ odd and negative for $k$ even. If we consider an amplituhedron for $k\neq n-4$, i.e. $k$ not maximal. The brackets $\braket{Yii+ijj+1}$ are the only brackets that have a fixed sign for all $Y$. This implies that the codimension one boundaries of the amplituhedron are a subset of the region described by the equation $\braket{Yii+1jj+1}=0$.

The bosonised superamplitude is obtained from the amplituhedron as its canonical form~\cite{Arkani-Hamed2017} which we will discuss further in section~\ref{oriented}.

\subsection{The amplituhedron and flipping number}

In~\cite{Arkani-Hamed2018} an equivalent, more direct definition of the amplituhedron was defined as a certain subspace of the set of oriented $k$-planes $Y$ in $k+4$ dimensions bounded by inequalities of the form $\braket{YZ_iZ_jZ_lZ_m}>0$, together with a further topological condition to be described, but importantly with no reference to the auxiliary positive matrix $C$ present in the original definition~\eqref{amplituhedron}.  

At tree-level  the alternative definition of the amplituhedron~\eqref{amplituhedron} is as the set
\begin{align}
	\sA_{n,k}:= \left\{Y \in Gr(k,k+4) \left| \begin{array}{ll}
		\braket{Yii+1jj+1}>0   &  1\leq i < j-1 \leq n-2\\
		\braket{Yii{+}11n}(-1)^{k}>0  & 1\leq i < n-1\\
		\{\langle Y 123i \rangle\} & \text{ has $k$ sign flips as $i=4,..,n$}
	\end{array}
	\right.
	\right\}\notag \\
	\text{for }Z\in Gr_{>}(k+4,n),
	\label{amplituhedron2}
\end{align}
That the two definitions~\eqref{amplituhedron} and~\eqref{amplituhedron2} are equivalent is proven for $m=1$ and $m=2$ \cite{Parisi:2021oql}, but still it is still conjectural for general $m$. 
Here the inequalities
$\braket{Yii+1jj+1}>0$ and $\braket{Yii+1n1} (-1)^{k+1}>0$
correspond to the locations of the proper poles of the amplitudes and are sometimes called  proper boundaries. The second set of constraints, is that the string $\{\langle Y 123i \rangle\}$ as $i$ ranges from 3 to $n$ must change sign exactly $k$ times, although the precise place where the sign changes is not important.
This is a purely topological condition and   $\braket{Y123i}=0$ will not be a physical boundary. 

This sign flip constraint  is clearly not manifestly cyclic. Cyclicity then demands that if the string $\{\braket{Y123i}\}$ has $k$ sign flips, then all the strings of the form  $\{\braket{Yjj{+}1j{+}2i}\}$ must have the same number of flips. Indeed, an even stronger statement can be proved. If the proper boundary inequalities hold, then all the strings of the form $\{\braket{Yj_1j_1{+}1j_2i}\}$ have the same number of flips as $i\neq j_1,j_1+1$ runs from $j_2{+}1$ to $j_2-1$%
~\cite{Arkani-Hamed2018}.

The loop amplituhedron can also be written in a similar form. The loop variables in the amplituhedron picture are represented by 2-planes $(AB)_i$ living in $Y^{\bot}$.  The loop amplituhedron $\sA_{n,k,l}$ is defined as the objects $\{Y,(AB)_{1},..,(AB)_l\}$, with $Y$ belonging to the tree level amplituhedron, and each $(AB)_i$ satisfying the following inequalities
\begin{align}
	\sA_{n,k,l}:= \left\{Y,(AB)_{1},.,(AB)_l \left| \begin{array}{ll}
		Y\in \sA_{n,k}&\\
		\braket{Y(AB)_jii+1}>0, &  \forall \ j,\ \forall i=1,.,n{-}1\\
		\braket{Y(AB)_j1n}(-1)^{k+1}>0 & \forall \ j\\
		\{\braket{Y(AB)_j1i} \} \qquad &\text{has } k{+}2 \text{ flips  as }i=2,..,n, \forall j \\
		\braket{(AB)_i(AB)_j}>0 & \forall i \neq j
	\end{array}
	\right.
	\right\}\notag \\
	\text{for }Z\in Gr_{>}(k+4,n),
	\label{amplituhedronloop}
\end{align}

\subsection{Amplituhedron-like geometries}
\label{amplike}

This new definition of the amplituhedron~\eqref{amplituhedron2} has the desirable feature of treating the proper boundaries and the other constraints separately, so we can modify the second while leaving the first the same. A natural generalization of these geometries is then to relax the constraint on the number of sign flips in~\eqref{amplituhedron2}. We thus define a tree-level amplituhedron-like geometry, $\sH_{n,k}^{(f)}$, by fixing the number of flips $f$.  To be consistent with cyclic or twisted cyclic invariance, the proper inequalities must be tweaked accordingly. Thus concretely we define  amplituhedron-like geometries
\begin{align}
	\sH_{n,k}^{(f)}:= \left\{Y \in Gr(k,k+4) \left| \begin{array}{ll}
		\braket{Yii+1jj+1}>0   &  1\leq i < j-1 \leq n-2\\
		\braket{Yii{+}11n}(-1)^{f}>0  & 1\leq i < n-1\\
		\{\langle Y 123i \rangle\} & \text{has $f$ sign flips as $i=4,..,n$}
	\end{array}
	\right.
	\right\}\notag \\
	\text{for }Z\in Gr_{+}(k+4,n),
	\label{amplituhedronlike}
\end{align}
In \cite{Arkani-Hamed2018} it was proven that  for a $k$-plane with convex $Z$s the maximal allowed number of flips is exactly $k$
so
\begin{align}
	0\leq f\leq k\ .
\end{align}
We can see that the  amplituhedron itself is then the case of an amplituhedron-like geometry with $f=k$, 
\begin{align}
	\sA_{n,k}=\sH_{n,k}^{(k)}\ .
\end{align}

The loop amplituhedron can also be generalised in a similar fashion. Here we allow for an arbitrary flipping  number, $f_j$, for each loop variable.
The generalization of the loop amplituhedron is given by a $Y$ belonging to a tree amplituhedron-like geometry and the loop variables satisfying
\begin{align}
	\sH_{n,k,l}^{(f;f_1,.,f_l)}:= \left\{Y,(AB)_{1},.,(AB)_l \left| \begin{array}{ll}
		Y\in \sH_{n,k}^{(f)}&\\
		\braket{Y(AB)_jii{+}1}>0, &  \forall \ j,\ \forall i=1,.,n{-}1\\
		\braket{Y(AB)_j1n}(-1)^{f_j}>0 & \forall \ j\\
		\{\braket{Y(AB)_j1i} \} \qquad &\text{has } f_j \text{ flips  as }i=2,..,n, \forall j \\
		\braket{Y(AB)_i(AB)_j}>0 & \forall i \neq j
	\end{array}
	\right.
	\right\}\notag \\
	\text{for }Z\in Gr_{>}(k+4,n)\,.
	\label{amplituhedronlikeloop}
\end{align}
The amplituhedron itself is then the case $f=k$ and $f_j=k+2$
\begin{align}
	\sA_{n,k,l}=\sH^{(k;k+2,k+2,..,k+2)}_{n,k,l}\ .
\end{align}
In the maximal $k$ case, $k=n-4$, $Z$ is a square $n\times n$ matrix and thus is always in $Gr_{>}(k+4,n)$ (or equivalently $Gr_{<}(k+4,n)$ if the determinant is negative).
Thus for much of what follows we will restrict to this case $k=n-4$.

Now we would like to see what are the possible values for the loop flipping numbers $f_j$.  If we project positive $Z$'s through a $k$-plane $Y$ with flipping number $f$, we obtain a configuration of $Z$'s on $Y^\perp$ that is defined by the brackets $\braket{ijkl}_Y=\braket{Yijkl}$.   The $\braket{ijkl}_Y$ satisfies the same inequalities as those of the N${}^f$MHV amplituhedron, $\sA_{n,f}$. In \cite{Arkani-Hamed2018} it is conjectured\footnote{The original formulation of conjecture is that given some $(m \times n)$ matrix of $Z$’s that satisfy the winding/flip criteria, we can always add $k$ more rows so that the resulting $(k {+} m) \times n$ matrix is positive.} that any $Z$ configuration $\braket{ijkl}$ with positive proper boundaries and flipping number equal to $f$ can be generated as a projection of positive $\tilde Z$s though an $f$-plane  $\tilde Y\in \sA_{n,f}$. This conjecture implies that for any $Y\in \sH_{n,k}^{(f)}$ there exists $\tilde{Y}\in \sA_{n,f}$, $\tilde Z\in Gr_>(f+m,n)$ and $\widetilde{AB} \in \tilde Y^\perp$ such that
\begin{align}
	\braket{ijkl}_Y=\braket{\tilde i\tilde j\tilde k\tilde l}_{\tilde{Y}}, \quad   \braket{ABij}_Y=\braket{\widetilde{AB}\tilde i\tilde j}_{\tilde{Y}}\quad   \forall \ i,j,k,l.
\end{align}
Therefore the sign flip string $\braket{YAB 1i}$ has the same constraints as the sign flip string $\braket{\tilde{Y} \widetilde{AB} \tilde 1\tilde i}$.
We know that the maximal flipping number for $k$-planes with positive $Z$s is $k$.
Here $(\tilde{Y} \widetilde{AB})$ is an $(f{+}2)$-plane and thus has maximal flipping number $f+2$. We can then conclude that the $\braket{YAB1i}$ flipping number must also be less than or equal to $f{+}2$. Moreover, the twisted cyclicity condition for $Y$ (second line of~\eqref{amplituhedronlike}) must be consistent with the twisted cyclicity condition for  each $(AB)_j$ (third line of~\eqref{amplituhedronlikeloop}).
We thus have the following restrictions on the loop flipping numbers $f_j$ in order to obtain a sensible geometry yielding a cyclic non-trivial canonical form
\begin{align}\label{flipcon}
	f_j\leq f+2, \qquad f_j=f \mod 2\ .
\end{align}
But there is a stronger  constraint which is easiest to see by considering the following  equivalence  map of geometries.

If we change the sign of alternate $Z$s, and all loop variables, we obtain a map between amplituhedron-like spaces with different flipping numbers,  $\sH_{n,n{-}4,l}^{(f;f_j)} \mapsto  \sH_{n,n{-}4,l}^{(n-4-f;\,n-2-f_j)}$. More concretely:
\begin{align}\label{dualgeometry}
	\sH_{n,n{-}4,l}^{(f;f_j)}\Big(Y,(AB)_j;Z_i\Big) =(-1)^{\lfloor \frac{n+1}2\rfloor} \sH_{n,n{-}4,l}^{(n-4-f;\,n-2-f_j)}\Big((-1)^{\lfloor \frac{n+1}2\rfloor}Y,-(AB)_j;Z_i(-1)^i\Big)\ ,
\end{align}
where an overall minus in front of $\sH$ indicates that we also reverse all the  inequalities (or equivalently send all $\braket{..}\mapsto -\braket{..}$.
This relation can be checked by just considering  the definitions on both sides. For example   the sign of every second element of the string $\braket{Y123i}$ is swapped under $Z_i \mapsto Z_i(-1)^i$. Thus every sign flip in the original space becomes a non sign flip and vice versa, and thus the flipping number $f \mapsto n{-}4{-}f$.

The canonical forms arising from the two geometries $\sH_{n,n{-}4,l}^{(f;f_j)}$ and $\sH_{n,n{-}4,l}^{(n-4-f;\,n-2-f_j)}$ are identical and we thus say that the geometries are ``equivalent''
\begin{align}\label{dualgeometry2}
	\sH_{n,n{-}4,l}^{(f;f_j)} \sim  \sH_{n,n{-}4,l}^{(n-4-f;\,n-2-f_j)}\ .
\end{align}
This equivalence then implies a much stronger bound on the allowed loop flipping numbers. We require $f_j \leq f+2$ but also for the dual geometry~\eqref{dualgeometry2} this means  $n-2-f_j \leq n-4-f+2$ ie $f \leq f_j$. Together with~\eqref{flipcon} we then see that each loop flipping number can only take 2 possible values
\begin{align}
	f_j = f \quad \text{or} \quad f+2\ .
\end{align}
With this in mind, 
we only need to keep track of the relative number of $f_j$s which are equal to $f+2$ and those which are equal to $f$. Finally it is also useful to define a geometry obtained by symmetrising over these variables. 
Thus we also define a loop amplituhedron-like geometry with just two superscripts, $f,l'$ where $l'$ is the number of loops with maximal flipping number $f+2$ 
\begin{align}\label{lprime}
	\sH_{n,n{-}4,l}^{(f;l')}&:= \bigcup_{\sigma\in S_l/(S_{l'}\times S_{l-l'})} \sH_{n,n{-}4,l}^{(f;\sigma(\overbrace{\scriptstyle f+2,..,f+2}^{l'},\overbrace{\scriptstyle f,..,f}^{l-l'}))}
	\ ,
\end{align}
where we take the union over all inequivalent choices of taking $l'$ loop variables to have maximal flipping number $f+2$ and the remaining ones minimal flipping number $f$.

\subsection{Conjecture: Amplituhedron-like  geometries give products}
\label{SquaredAmplituhedron}

Having defined a natural generalisation of the amplituhedron, the amplituhedron-like geometries, we now discuss what they correspond to physically. First at tree level, focusing on the maximal $k=n-4$, there are  $k+1$ amplituhedron-like geometries  $\sH_{n,k}^{(f)}$, $f=0,..,k$ with $f$ equivalent to $k-f$ through~\eqref{dualgeometry2}. This perfectly mimics the possible products of two amplitudes of total Grassmann degree $k=n-4$, $\cA_{n,k'}\cA_{n,n-4-k'}$. We  conjecture that the canonical form $H$ (in fact the oriented canonical form to be defined in the next section) of an amplituhedron-like geometry $\sH$ gives the star product (see~\eqref{diag}) of superamplitudes
\begin{align}
	\text{main conjecture  (tree-level):   } \quad \boxed{	H^{(f)}_{n,n-4} = A_{n,f}* A_{n,n-f-4}\,.}
	\label{conjecture1}
\end{align}

Note that in the case of maximal flipping number,  $f=n-4$, this conjecture collapses to the standard amplituhedron conjecture (recalling that $A_{n,0}=1$).
We will define the canonical form in the next section and then in the following section describe the various proofs and checks giving evidence for this conjecture which we have performed.

The amplituhedron-like geometries at loop level depend also on the flipping number of the loop variables, $l'$ (see~\eqref{lprime}). 
We thus generalize~\eqref{conjecture1} to loop level and conjecture that for $k=n-4$ the canonical form of the loop amplituhedron-like geometry with $l'$ loops having maximal flipping number and $l-l'$ loops minimal is 
\begin{align}\label{conjecture2a}
	H_{n,n{-}4,l}^{(f;\overbrace{\scriptstyle f+2,..,f+2}^{l'},\overbrace{\scriptstyle f,..,f}^{l-l'})}= A_{n,f,l'}(AB_1,\cdots,AB_{l'}) * A_{n,n-f-4,l-l'}(AB_{l'+1},\cdots,AB_{l})
\end{align}
where the loop variables with maximal flipping number $f+2$ belong to the first factor $A_{n,k',l'}$ and the remaining loop variables to $A_{n,n-k'-4,l-l'}$.
By summing over inequivalent permutations of the loops we then obtain 
\begin{align}
	H_{n,n-4,l}^{(k',l')}= \sum_{\sigma \in {S_l}/({S_{l'}\times S_{l-l'}})}\,A_{n,k',l'}((AB)_{\sigma(i)}) * A_{n,n-k'-4,l-l'}((AB)_{\sigma(i)})\,,
	\label{conjecture2b}
\end{align}
suppressing the explicit distribution of loop variables this can be written in the more   compact form 
\begin{align}
	\text{main conjecture (loop)-level  }: \quad \boxed{H_{n,n-4,l}^{(k',l')}= \begin{psmallmatrix}l\\l'\end{psmallmatrix}\,A_{n,k',l'} * A_{n,n-k'-4,l-l'}\,.}
	\label{conjecture2}
\end{align}

One can see that \eqref{conjecture2} is consistent with the duality~\eqref{dualgeometry2} and it's trivially true  for the case $k'=n-4, l'=l$ which collapses to the standard amplituhedron conjecture for the anti-MHV loop level amplitude
\begin{align}
	H_{n,n-4,l}^{(n-4,l)}=  A_{n,n-4,l}*A_{n,0,0}= \, A_{n,n-4,l}.
\end{align}

Last but not least, the conjecture \eqref{conjecture2} is consistent with the   squared amplituhedron conjecture. In particular we define the squared amplituhedron as the union of two geometries defined by physical inequalities only, ie with no topological winding condition. The two geometries are distinguished purely by their properties under cyclicity: twisted or untwisted
\begin{align}\label{sqamp}
	\text{Squared amplituhedron: }\qquad   \sH_{n,n-4,l}:=\sH^+_{n,n-4,l} \cup \sH^-_{n,n-4,l}
\end{align}
with
\begin{align}\label{sqamplooppm}
	\sH^\pm_{n,k,l}:= \left\{Y,(AB)_{1},..,(AB)_l \left| \begin{array}{ll} \braket{Yii+1jj+1}>0   &  1\leq i < j-1 \leq n-2\\
		\pm \braket{Yii{+}11n}> 0  & 1\leq i < n-1\\
		\braket{Y(AB)_jii{+}1}>0 &  \forall \ j,\ \forall i=1,..,n{-}1\\
		\pm \braket{Y(AB)_j1n}> 0 & \forall \ j\\
		\braket{(AB)_i(AB)_j}>0 & \forall i \neq j
	\end{array}
	\right.
	\right\}\notag \\
	\text{for }Z\in Gr_{>}(k+4,n)\,.
\end{align}
The amplituhedron-like geometries are clearly subsets of $\sH^\pm_{n,k,l}$, Furthermore the union of all even/odd flipping numbered amplituhedron-like geometries clearly gives $\sH_{n,n-4,l}^\pm$ and the union of all amplituhedron-like geometries gives the squared amplituhedron $\sH_{n,n-4,l}$
\begin{align}\label{sqaunion}
	\sH_{n,n-4,l}=\bigcup_{f,l'} \sH^{(f,l')}_{n,n-4,l} \ .
\end{align}
Comparing with the expansion of the amplitude squared~\eqref{loopsa} into precisely the same products we get from the amplituhedron-like geometries,  
it is thus natural to conclude that the canonical form of $\sH_{n,n-4,l}$ is the square of the amplitude. However one has to be a bit more careful. As we describe in the next section, the canonical form is defined very rigidly, only for very specific types of geometrical spaces. As  such the standard canonical form of $\sH_{n,k,l}$ is not defined. However as we will see a very natural extension of the canonical form, the ``oriented canonical form'' can indeed be defined for $\sH_{n,k,l}$. 
To prove that the oriented canonical form of the squared amplituhedron is equal to the square of the superamplitude starting from \eqref{conjecture2} we would need to prove that all the amplituhedron-like geometries have the right orientation, so that the product of amplitudes sum with the right signs. We don't know how to prove this in general but we have checked explicitly that tree level computations for $k\leq3$ and loop computations for $n=4,l=2$ and $n=5,l=1$ are consistent with this conjecture \cite{Eden2017}.

\subsection{General $m$ amplituhedron-like geometries}

As was already pointed out in the original amplituhedron  paper~\cite{Arkani-Hamed:2016byb}, the definition of the {tree} amplituhedron can be generalized to arbitrary twistor dimension, $m$. The same can clearly be done for the tree amplituhedron-like geometries. In the generalisation we have $Y \in Gr(k,k+m)$ rather than $Gr(k,k+4)$ and  the $Z$s live in $k+m$ dimensions rather than $k+4$. The defining inequalities of the amplituhedron-like geometry, $A_{n,k}^{(f)}$, are then similar to the $m=4$ case~\eqref{amplituhedronlike}) with the following modifications.
The sign flip string for generic $m$ reads (compare with~\eqref{amplituhedronlike})
\begin{align}
	\{\braket{Y\,123\cdots (m{-}1)i}\}\qquad \text{has $f$ sign flips as $i=m,m{+}1,..,n$} \ ,
\end{align}
and the physical inequalities  read
\begin{align}
	\left.
	\begin{array}{rcl}
		\braket{(i_1 i_1+1) \cdots i_{\frac{m}{2}}(i_{\frac{m}{2}}+1}&>&0 \\ \braket{i_1 i_1+1 \cdots i_{\frac{m}{2}}(i_{\frac{m}{2}}+1)1n}(-1)^{f}&>&0 
	\end{array}
	\right\}
	&\text{ for }m \text{ even}, \nonumber \\
	\left.
	\begin{array}{rcl}
		\braket{1i_1 i_1+1 \cdots i_{\frac{m-1}{2}}(i_{\frac{m-1}{2}}+1)}&>&0   \\ 
		\braket{i_1 i_1+1 \cdots i_{\frac{m-1}{2}}(i_{\frac{m-1}{2}}+1)n}(-1)^{f}&>&0
	\end{array}
	\right\}
	&\text{ for }m \text{ odd}.
	\label{generalizedAmp}
\end{align}

Much of the analysis that we did for $m=4$ also applies to general $m$. In particular the duality relation \eqref{dualgeometry2} becomes
\begin{align} \label{dualgeneralgeometry}
	\sH_{n,n{-}m}^{(f)} \sim \sH_{n,n{-}m}^{(n-m-f)}\ ,
\end{align}
and we conjecture  that the canonical form of the maximal generalised amplituhedron-like geometries are products in a similar way to \eqref{conjecture1} 
\begin{align}\label{conjecture1gen}
	H^{(f)}_{n,n-m} = A_{n,f}*A_{n,n-m-f}
\end{align}
where $A_{n,f}:= H_{n,f}^{(f)}$, the canonical form of the standard (but generalised $m$) amplituhedron.

\subsection{Amplituhedron-like geometries: alternative definition} 
\label{altdef}
The original definition of the amplituhedron was given as the projection of the positive Grassmannian $Gr_>(k,k+n)$ through positive $Z$s onto $Gr(k,k+m)$~\eqref{amplituhedron}. We have then defined amplituhedron-like geometries as generalisations of the alternative flipping number definition of the amplituhedron. It is then interesting to see if there is an alternative definition of the amplituhedron-like geometries which generalises the original definition of the amplituhedron. Here we propose precisely such an equivalent definition for the maximal case. 
We propose that the maximal $k=n-m$  (generalised) amplituhedron-like geometry,  $\sH_{n,n-m}^{(f)}$, can be written as the  projection of the positive Grassmannian $Gr_{>}(f,n)$ and the alternating positive Grassmannian $\text{alt}(Gr_{>})(n-m-f,n)$ through the positive $Z$s onto $Gr(k,k+m)$:
\small
\begin{align}\label{GrassmannianProduct}
	\sH_{n,n-m}^{(f);\text{alt}} :=\begin{cases} Y=\begin{pmatrix}C_1\\C_2\end{pmatrix}\cdot Z\ |\ C_1\in Gr_{>}(f,n) \ \land  \  C_2\in \text{alt}(Gr_{>})(n-m-f,n)\ &\text{for $g_{n,f}$ even}\\
		Y=\begin{pmatrix}C_1\\C_2\end{pmatrix}\cdot Z\ |\ C_1\in Gr_{<}(f,n) \ \land  \  C_2\in \text{alt}(Gr_{>})(n-m-f,n)\ &\text{for $g_{n,f}$ odd}
	\end{cases}
\end{align}
\normalsize
where $g_{n,f}:=	\lfloor\frac{n-f}{2}\rfloor+(n{-}f)n$.
Here the alternating positive Grassmannian, alt(Gr)$_{>}(k,n)$, is defined as the image of Gr$_>(k,n)$ under the transformation which flips the sign of the odd columns. 
We will give evidence for the equivalence of this definition of the amplituhedron-like geometry with the flipping number definition~\eqref{amplituhedronlike} in section~\ref{AmpAsProd}.

Notice that for maximal $f= n-m$ this definition coincides with the original amplituhedron. However for general $f$ the geometry splits into two copies of the amplituhedron. This product geometry  manifests the conjecture that the canonical form of this geometry gives the product of the corresponding amplitudes~\eqref{conjecture1gen}. 

This definition naturally extends to loops. For the amplituhedron we have that each loop can be parametrised using an auxiliary $2\times n$ matrix $D_i$ as $AB_i=D_i\cdot Z$ with the condition
$\begin{pmatrix}C\\D_i\end{pmatrix}\in Gr_>(k+2,n)$, which corresponds to the one loop constraints and $\begin{pmatrix}C\\D_i\\D_j\end{pmatrix}\in Gr_>(k+4,n)$ which corresponds to mutual positivity~\cite{Arkani-Hamed2014}. In analogy to the loop amplituhedron  we can define the loop amplituhedron-like geometry (setting here $Z={\mathbb{1}}$)
\begin{align}\label{loopGrassmannianProduct}
	\sH_{n,n{-}4,l}^{(f;\overbrace{\scriptstyle f+2,..,f+2}^{l'},\overbrace{\scriptstyle f,..,f}^{l-l'})}:=\begin{cases}   \begin{pmatrix}C_1\\D_i\end{pmatrix}\in Gr_>(f+2,n) \quad \forall \quad i\leq l'\\   \begin{pmatrix}C_2\\D_i\end{pmatrix}\in \text{alt}(Gr_>)(n-f-2,n) \quad \forall \quad i> l' \\ \det \begin{pmatrix} C\\D_i\\D_j\end{pmatrix} >0\quad \forall \quad i\neq j
	\end{cases}
\end{align}
where the tree level condition \eqref{GrassmannianProduct} is understood.

\section{ The globally oriented canonical form}
\label{oriented}

As discussed in section~\ref{maxres} Even though the bosonised squared superamplitude has only dlog singularities it misses one crucial property for it to be represented as the canonical form of a positive geometry: its  maximal residues are not normalizable to $\pm1,0$. Nevertheless, in \cite{Eden2017}  the square of the superamplitude was obtained from the geometry of the maximal squared amplituhedron using   the CAD (Cylindrical Algebraic Decomposition, see  \ref{CAD}). This happens because the squared amplituhedron is actually not a positive geometry, but its interior is made of the union of almost disconnected components, that is disconnected regions that do not share a codimension 1 boundary but may share lower dimension boundaries. The disconnected components themselves are positive geometries but the union is not. For these types of geometries the CAD algorithm doesn't necessarily output the canonical form (indeed in most cases the canonical form would not be defined) but it does give a unique form which we will call the {\em globally oriented canonical form} or {\em oriented canonical} form for short. 

A key point is that the CAD algorithm is defined on the {\em oriented} Grassmannian, which is an orientable manifold,  rather than the Grassmannian itself which is  generally not orientable. On orientable manifolds there exists a global notion of ``positive" orientation.  We can use this notion to  fix the relative orientation of disconnected geometries by imposing that they are all  positively oriented. For this choice of orientation, residues on shared codimension 1 boundaries automatically cancel, but residues on lower codimension boundaries can sum. We will start by recalling the definition of the canonical form and then we will use it to define the oriented canonical form.\\

\subsection{The canonical form}
\label{semialg}

In~\cite{Arkani-Hamed2017} positive geometries and their  canonical forms were defined. A positive geometry is by definition a geometry that has a canonical form, and both the concept of positive geometry as well as its canonical form are defined recursively. A $D$-dimensional positive geometry is defined as the pair $(X,X_{\geq 0})$ possessing a  canonical form $\omega(X,X_{\geq0})$ satisfying the following conditions
\begin{itemize}
	\item $X$ is a  complex projective algebraic variety of complex dimension $D$, known as the embedding space. In practice for our application the  algebraic variety will be a Grassmannian but the definition is given in this more general setting.
	\item $X_{\geq 0}$, is a closed,   oriented,  $D$-dimensional semi-algebraic subset of $ X(\mathbb{R})$, the real slice of $X$.%
	\item There is a unique top form $\omega(X,X_{\geq 0})$ called the canonical form.
	\item Every boundary component $(C,C_{\geq0})$ is itself a positive geometry of dimension $D-1$.
	\item The canonical form has no singularities inside $X_{\geq 0}$, but has simple poles on the boundary.  The recursive step is then that the residue of the canonical form on each  boundary component   is  equal to the canonical form of the boundary component itself:  $$\text{Res}_{C}\left(\omega(X,X_{\geq 0})\right)=\omega(C,C_{\geq0})\,.$$
	\item The recursion is initiated by defining   0-dimensional positive geometries, for which $X_{\geq 0}$ is just a single point and $\omega(X,X_{\geq 0})=\pm 1$ depending on the orientation.  
\end{itemize}

In the following we will follow convention and often simplify notation and refer to the positive geometry simply by $X_{\geq 0}$ instead of $(X,X_{\geq 0})$.

Note that $X_{\geq 0}$ is defined as  a semi-algebraic subset of $X(\mathbb{R})$ which itself is a subset of $\mathbb{P}^{n}$. A semi-algebraic set is defined by a set of homogeneous real polynomial equations, $p(x)=0$, and inequalities, $q(x)>0$. Now inequalities $q(x)>0$ are problematic in projective spaces since homogeneous coordinates are invariant under $x\rightarrow -x$ which may flip  the sign and change the inequality. For this reason the prescription is to first define the region in $\mathbb{R}^{n+1}\slash \{0\}$ and then project onto $\mathbb{P}^n$ to obtain $X_{\geq 0}$. 
However we could equally project instead onto  $\mathbb{R}^{n+1}/\mathbb{R}^+$, oriented projective space instead of projective space itself. Nothing in the definition of the canonical form or positive geometries appears to rely on being define in projective space rather than oriented projective space and we will take this point of view later when defining the globally oriented canonical form.

\subsection{The union of positive geometries}

Having reviewed the definition of positive geometries and their associated canonical forms, we now consider the union of positive geometries for various cases. A similar discussion can be found in~\cite{Damgaard:2021qbi}.

As was already discussed in~\cite{Arkani-Hamed2017}, the union of two completely disjoint positive geometries  $X_1$,  $X_2$ is itself a positive geometry and the canonical form is simply the sum of the two canonical forms: $\omega (X_1 \cup X_2)= \omega (X_1) + \omega(X_2)$. Since the geometries are disjoint, and the standard canonical form requires no  concept of global orientation, each disjoint piece can come in either orientation the signs of either term depend on this choice.

A more interesting case to consider is that of two positive geometries $X_1,X_2$ which only overlap on their boundary. Firstly consider the case where they share a codimension 1 boundary. The union can only form a positive geometry if the orientations of $X_1$ and $X_2$ agree. If the orientations  do agree then the canonical forms along the common boundary of $X_1$ and $X_2$  will cancel (as it must for this to be a positive geometry as this will lie in the interior of the union). A simple example is that of two triangles sharing an edge:
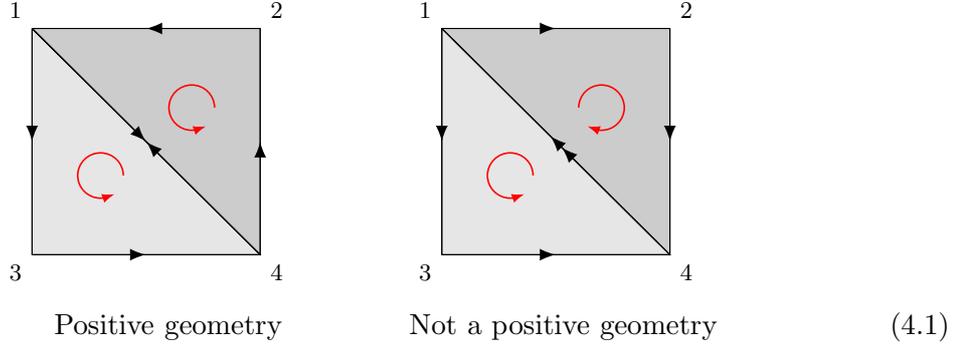
\begin{align}
	\label{ex1}
	\begin{tikzpicture}[x=1.5cm,y=1.5cm,decoration={markings, 
			mark= at position 0.5 with {\pgftransformscale{1.5}\arrow{latex}}}]
		\filldraw[fill=gray!20] (-1,1) node[above left]{\footnotesize 1}--(1,-1)node[below right]{\footnotesize 4}--(-1,-1)node[below left]{\footnotesize 3}--(-1,1);
		\filldraw[fill=gray!40] (-1,1)--(1,-1)--(1,1)node[above right]{\footnotesize 2}--(-1,1);
		\draw[postaction={decorate}] (1,-1)--(-1,1);
		\draw[postaction={decorate}] (-1,1)--(-1,-1);
		\draw[postaction={decorate}] (-1,-1)--(1,-1);
		\draw[postaction={decorate}] (-1,1)--(1,-1);
		\draw[postaction={decorate}] (1,-1)--(1,1);
		\draw[postaction={decorate}] (1,1)--(-1,1);
		\draw[->,>=latex,semithick,red] (.6,.3) arc[radius=.2, start angle=0, end angle=300]-- ++(20:1pt);
		\draw[->,>=latex,semithick,red] (-.2,-.3) arc[radius=.2, start angle=0, end angle=300]-- ++(20:1pt);
	\end{tikzpicture}\qquad \qquad &
	\begin{tikzpicture}[x=1.5cm,y=1.5cm,decoration={markings, 
			mark= at position 0.5 with {\pgftransformscale{1.5}\arrow{latex}}}]
		\filldraw[fill=gray!20] (-1,1) node[above left]{\footnotesize 1}--(1,-1)node[below right]{\footnotesize 4}--(-1,-1)node[below left]{\footnotesize 3}--(-1,1);
		\filldraw[fill=gray!40] (-1,1)--(1,-1)--(1,1)node[above right]{\footnotesize 2}--(-1,1);
		\draw[postaction={decorate}] (1,-1)--(-.9,.9);
		\draw[postaction={decorate}] (-1,1)--(-1,-1);
		\draw[postaction={decorate}] (-1,-1)--(1,-1);
		\draw[postaction={decorate}] (.9,-.9)--(-1,1);
		\draw[postaction={decorate}] (1,1)--(1,-1);
		\draw[postaction={decorate}] (-1,1)--(1,1);
		\draw[->,>=latex,semithick,red] (.2,.3) arc[radius=.2, start angle=180, end angle=-120]-- ++(160:1pt);
		\draw[->,>=latex,semithick,red] (-.2,-.3) arc[radius=.2, start angle=0, end angle=300]-- ++(20:1pt);
	\end{tikzpicture}\notag\\
	\text{Positive geometry }\qquad \qquad  & \text{Not a positive geometry}
\end{align}

Now consider a union  of two positive geometries sharing a boundary of lower dimension, for example two triangles touching at a vertex: 
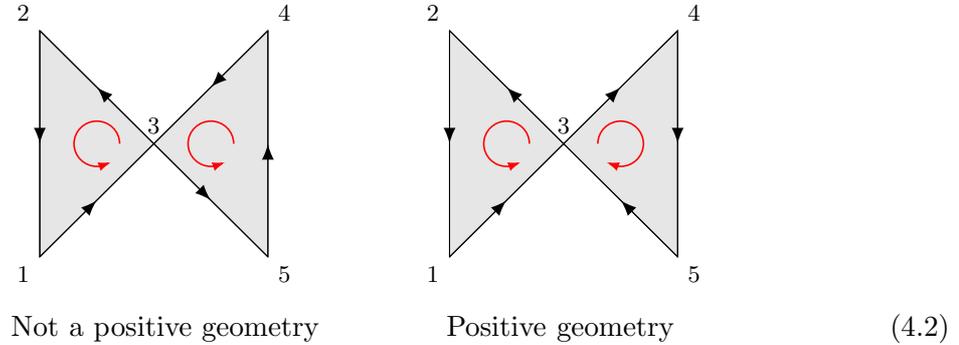
\begin{align}
	\label{ex2}
	\begin{tikzpicture}[x=1.5cm,y=1.5cm,decoration={markings, 
			mark= at position 0.5 with {\pgftransformscale{1.5}\arrow{latex}}}
		] 
		\filldraw[fill=gray!20] (0,0) node[above]{\footnotesize 3}--(-1,1) node[above left]{\footnotesize 2}--(-1,-1)node[below left]{\footnotesize 1}--(0,0);
		\draw[postaction={decorate}] (0,0)--(-1,1);
		\draw[postaction={decorate}] (-1,1)--(-1,-1);
		\draw[postaction={decorate}] (-1,-1)--(0,0);
		\filldraw[fill=gray!20] (0,0)--(1,-1)node[below right]{\footnotesize 5}--(1,1)node[above right]{\footnotesize 4}--(0,0);
		\draw[postaction={decorate}] (0,0)--(1,-1);
		\draw[postaction={decorate}] (1,-1)--(1,1);
		\draw[postaction={decorate}] (1,1)--(0,0);
		\draw[->,>=latex,semithick,red] (.7,0) arc[radius=.2, start angle=0, end angle=300]-- ++(20:1pt);
		\draw[->,>=latex,semithick,red] (-.3,0) arc[radius=.2, start angle=0, end angle=300]-- ++(20:1pt);
	\end{tikzpicture}
	\qquad \qquad &
	\begin{tikzpicture}[x=1.5cm,y=1.5cm,decoration={markings, 
			mark= at position 0.5 with {\pgftransformscale{1.5}\arrow{latex}}}
		] 
		\filldraw[fill=gray!20] (0,0) node[above]{\footnotesize 3}--(-1,1) node[above left]{\footnotesize 2}--(-1,-1)node[below left]{\footnotesize 1}--(0,0);
		\draw[postaction={decorate}] (0,0)--(-1,1);
		\draw[postaction={decorate}] (-1,1)--(-1,-1);
		\draw[postaction={decorate}] (-1,-1)--(0,0);
		\filldraw[fill=gray!20] (0,0)--(1,-1)node[below right]{\footnotesize 5}--(1,1)node[above right]{\footnotesize 4}--(0,0);
		\draw[postaction={decorate}] (1,-1)--(0,0);
		\draw[postaction={decorate}] (1,1)--(1,-1);
		\draw[postaction={decorate}] (0,0)--(1,1);
		\draw[->,>=latex,semithick,red] (.3,0) arc[radius=.2, start angle=180, end angle=-120]-- ++(160:1pt);
		\draw[->,>=latex,semithick,red] (-.3,0) arc[radius=.2, start angle=0, end angle=300]-- ++(20:1pt);
	\end{tikzpicture}\notag\\
	\text{Not a positive geometry } \qquad  \quad & \quad \text{Positive geometry}
\end{align}
Here the case where the orientations agree is {\em not} a positive geometry whereas the case where they disagree is. 
To see this let's consider the canonical forms in the two cases. The canonical form of a triangle $\{i,j,k\}$  with standard orientation  is 
\begin{align}
	\label{ex3}
	\omega_{ijk}= \frac{\braket{Yd^2Y}\braket{ijk}}{\braket{Yij}\braket{Yjk}\braket{Yki}}
\end{align}
and thus the canonical form of the union of the two triangles, if it exists, will be given by $\omega_{123}+\omega_{345}$ or $\omega_{123}-\omega_{345}$ in the two cases respectively.  In  the first case the double residue corresponding to the residue at vertex 3 is%
\footnote{Note that $\braket{Y23}\to 0 \Leftrightarrow \braket{Y35}\to 0$.}
\begin{align}
	\text{Res}_{\braket{Y13}\to0}\left(	    \text{Res}_{\braket{Y23}\to0}\left( \omega_{123}+\omega_{345}\right)\right)=-2,
\end{align}
which is different from $\pm 1,0$. In the second case instead the residue is simply zero. Double residues at the other points are equal to  $\pm 1,0$ since only one triangle at a time will contribute.  Thus only the second geometry  is a positive geometry. Note the difference with the previous case where the orientations had to agree for a positive geometry, here instead they have to disagree for it to be a positive geometry!

Finally consider the union of the two triangles we considered before with the addition of a rectangle below
\begin{align}
	&\qquad\begin{tikzpicture}[x=1.5cm,y=1.5cm]
		\filldraw[fill=gray!20] (0,0) node[above]{\footnotesize 3}--(-1,1) node[above left]{\footnotesize 2}--(-1,-1)node[below left]{\footnotesize 1}--(0,0);
		\filldraw[fill=gray!20] (0,0)--(1,-1)node[below right]{\footnotesize 5}--(1,1)node[above right]{\footnotesize 4}--(0,0);
		\filldraw[fill=gray!20] (-1,-1)--(1,-1)--(1,-2)node[below right]{\footnotesize 6}--(-1,-2)node[below left]{\footnotesize 7}--(-1,-1);
	\end{tikzpicture}\notag \\
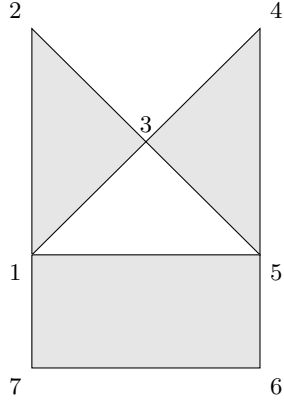
	& \qquad\text{Not a positive geometry} \notag\\
	&\text{(for any choice of orientations)}
\end{align}	     	 
We concluded before that to avoid the $\pm 2$ residue at vertex $3$ we need the two triangles to have opposite orientation. Similarly, to have residue equal to zero on $Z_1$ we need the rectangle to have the same orientation as the triangle $\{1,2,3\}$. However  we also need  triangle $\{3,4,5\}$ to have the same orientation as the rectangle to get the right maximal residue on $Z_5$, but opposite orientation to the triangle $\{123\}$ to have the right residue on $Z_3$. These constraints are clearly incompatible and therefore this geometry is not a positive geometry and does not possess a canonical form.

\subsection{The globally oriented canonical form}

The above examples illustrate the very precise nature of the  definition of a positive geometry and motivate the investigation of generalisations of  this definition to include some of the above unions. 
As we have seen,  each separate positive geometry can have its own orientation and only if these are chosen appropriately do we still obtain a positive geometry when they touch, and indeed this is by no means always possible as illustrated in the last example.
The essential problem is that as soon as positive geometries touch, there is the  possibility of the maximal residues at intersecting points summing to values differing from $\pm 1,0$. 
Now, as discussed in section~\ref{maxres}, the square of the amplituhedron does have maximal residues which can differ from $\pm 1,0$.
This motivates us therefore to consider more general geometries and an extension of the definition of the canonical form  to allow for such cases.
Firstly it seems appropriate to fix an unambiguous global orientation. One very simple way to do this if  $X$ itself is orientable is to simply inherit the orientation from $X$. The problem then is that $X$ is not always orientable. 
However, it is also the case that, as mentioned at the end of section~\ref{semialg}, a positive geometry can be  defined in the oriented projective space $\mathbb{R}^{n+1}/\mathbb{R}^+$
which is always orientable (it's a double cover of $P^n$ and equivalent to the sphere $S^n$). 

We thus formally define a globally oriented canonical form $\Omega$ and the corresponding spaces which possess one, the pair $(X,X_{\geq 0})$ as follows:
\begin{itemize}
	\item $X$ is an irreducible complex projective variety.
	\item The  double cover of $X(\mathbb{R})$ is orientable. This is always true for the case of direct interest where $X(\mathbb{R})=Gr(k,k+m)$ and the double cover is  $\tilde Gr(k,k+m)$, the oriented Grassmannian, which is orientable.%
	\footnote{\label{f5} A global  orientation on the oriented Grassmannian $\tilde Gr(k,k+m)$ can be defined via the global top form $$ \prod_{i=1}^{k+m}\frac{\braket{Y d^m Y_i}}{\sqrt{|\det(Y Y^T)|}}\,.$$
		This transforms by a factor $\det(G)^{k+m}/|\det(G)|^{k+m}$ under $Y\to G Y$ and is thus well-defined on the oriented Grassmannian (on which $\det(G)>0$) but not on the Grassmannian itself for $k+m$ odd.  For the loop level amplituhedron there is a  similar expression by including for each loop variable  
		$$ \frac{\braket{A_j B_j d^2 A_j}\braket{A_j B_j d^2 B_j}}{(|A|^2|B|^2-(A.B)^2)^2}\,.$$
		In both tree level and loop level cases one sees that it is the sign of the numerator  (eg   $\prod_{i}
		\braket{Y d^m Y_i}$) which determines the orientation. 
	}
	\item $X_{\geq0}$ is a closed  $D$-dimensional semi-algebraic subset of this double cover of $X(\mathbb{R})$.
	\item $X_{\geq0}$ is the union  of a set of positive geometries $X_{\geq0}= \bigcup_i X^{(i)}_{\geq 0}$ whose interiors are connected and mutually disjoint, so
	for any pair $X^{(i)}_{> 0} \cap X^{(j)}_{> 0}=\emptyset$. (Here we take the positive geometries to be defined in the double cover rather than in $X(\mathbb{R})$ directly,  as discussed at the end of section~\ref{semialg}.)
	\item The orientations of the positive geometries $X^{(i)}_{\geq 0}$ are inherited from that of (the double cover of) $X(\mathbb{R})$. 
	\item
	The {\em globally oriented canonical form} $\Omega$ of $(X,X_{\geq 0})$ is then simply defined to be  the sum of the canonical forms $\omega$ of the positive geometries $(X,X^{(i)}_{\geq 0})$:
	\begin{align}
		\Omega(X_{\geq 0}) = \sum_i \omega(X^{(i)}_{\geq 0})\ .
	\end{align}
\end{itemize}
We believe this gives a unique definition. In other words if $X_{\geq 0}$ can be described as a union of positive geometries in two different ways, $X_{\geq0}= \bigcup_i X^{(i)}_{\geq 0}= \bigcup_j X'^{(j)}_{\geq 0}$, the resulting sums of canonical forms should be equal, $\sum_i \omega( X^{(i)}_{\geq 0})=\sum_j \omega( X'^{(j)}_{\geq 0})$. This essentially follows from similar arguments to those establishing triangulation independence of the canonical forms of positive geometries (see section 3 of~\cite{Arkani-Hamed2017}).

As a consequence of this definition, the oriented canonical form of a positive geometry with connected interior is equal to the canonical form. However, in general,  the maximal residues of an oriented canonical form can be different from $\pm1,0$.

It is interesting to revisit the examples of the previous subsection. The example of two positive geometries intersecting on a codimension 1 boundary~\eqref{ex1} with the same orientation gives a positive geometry and the canonical and oriented canonical forms agree. The example of two triangles touching at a vertex~\eqref{ex2} shows a difference between the two case. The first case has an oriented canonical (but not a canonical form)  whereas the second case does not have an oriented canonical form (but does have a canonical form).
Finally in the third example~\eqref{ex3} we saw that for no choice of orientations on the three shapes could this be a positive geometry and have a canonical form. 
Nevertheless, it possesses an oriented canonical form which is the sum of the positively oriented canonical forms of the two triangles and the rectangle. Indicating with $\Omega$ the oriented canonical form we have
\begin{align}
	{\Omega}= \omega_{123}+\omega_{345}+\omega_{157}+\omega_{567}.
\end{align}
Let's see now an algorithm that can be used in general to compute the oriented canonical form of amplituhedron-like geometries.

\subsection{Oriented canonical form made simple: The CAD algorithm}
\label{CAD}

The definition of the canonical form and in turn the oriented canonical form which is given in terms of it is quite intricate
and finding an algorithm to systematically compute the canonical form of a generic positive geometry is still an open problem.  Indeed it is far from obvious if any given geometry even has  a canonical form or indeed an oriented canonical form.  The general strategy is to triangulate the positive geometry into a set of regions for which the canonical form is known and then take the sum of the canonical forms with the appropriate relative signs.

The situation simplifies enormously for the oriented canonical form  however, at least  if the geometry $X_{\geq 0}$ is defined only by linear inequalities, due to the following:
\vskip15pt
\noindent\fbox{%
	\parbox{\textwidth}{%
		For any semi-algebraic space $X_{\geq 0}$ (a subset of the oriented double cover of an irreducible complex projective variety) for which  we  can choose coordinates with respect to  which $X_{\geq 0}$ is defined by multi-linear inequalities, then:
		\begin{enumerate}
			\item $X_{\geq 0}$ has a global oriented canonical form.
			\item There is a simple algorithm for computing the oriented canonical form using Cylindrical Algebraic Decomposition (CAD)~\cite{10.1145/1093390.1093393}.
		\end{enumerate}
	}%
}
\vskip15pt

We will now discuss how to use the CAD algorithm 
to compute canonical forms, as described in \cite{Eden2017}. Then we will show that this algorithm gives the canonical form for connected geometries and more  generally gives the oriented canonical form. For simplicity we will describe this in the context of  a subset of the oriented  Grassmannian  $\widetilde{Gr}(k,k{+}m)$ (see discussion below~\eqref{amplituhedron}),  but the generalisation to more general cases such as including $l$ planes at loop level or simply a general complex variety satisfying the conditions stated previously should be straightforward. 

Suppose we what to compute the canonical form of a geometry  $ \mathcal{T}$ in $\widetilde{Gr}(k,k{+}m)$ .
The starting point is to give numeric coordinates to the $Z$ external data and to parametrise $Y\in \widetilde{Gr}(k,k{+}m)$ as a $k\times (k+m)$ matrix depending on $k\times m$ variables $\{x_1,\cdots x_{km}\}$.  Giving coordinates to $Y$ corresponds to creating a map
\begin{align}
	\phi: \mathbb{R}^{km} \qquad \rightarrow \qquad \widetilde{Gr}(k,k+m),
\end{align}
that in general will cover  half of the oriented Grassmannian.  If the geometry is not all covered by a single coordinate patch, one needs to first triangulate the region into sub-regions each of which can be covered by a single patch. Then, compute the canonical form of each sub-region, using CAD, each in their own coordinates. Finally one needs to sum the resulting canonical forms. To sum the canonical forms one needs to write them in the same coordinates however. Since rational forms are holomorphic functions, once known on a coordinate chart, they  can be analytically continued in a unique way to the whole Grassmannian and therefore to any other chart. To analytically continue a canonical form to the whole Grassmannian is equivalent to computing its  covariant form.   We give an example of this summing procedure at the end of this section.

Given an  ordering of the coordinates $\{x_1,\cdots x_{km}\}$ on some patch, then a cylindrical decomposition of a subset of $\mathbb{R}^{km}$ describes it as a union of regions $\mathcal{R}_i$ defined by inequalities of the form
\begin{align}
	\mathcal{R}_{i}:= \{x_1,\cdots,x_{km}\}\in \mathbb{R}^{mk} \quad \text{st} \qquad \begin{cases} \hfil  a_1<x_1<b_1  \\  \hfil  a_2(x_1)<x_2<b_2(x_1)\\
		\hfil   \cdots \\
		a_{km}(x_1,\cdots,x_{km-1})<x_{km}<b_{km}(x_1,\cdots,x_{km-1})
	\end{cases}
	\label{easyregion}
\end{align}
for some functions $a_j(x_1,..,x_{j-1})$. Note that this is just  the procedure one would take for converting a multiple integral over the region $\cal T$ into a sum of repeated single integrals, the inequalities~\eqref{easyregion} being the limits of the resulting integrals.

Thus if the coordinates $\phi$ cover the whole of $\mathcal{T}$,  the cylindrical decomposition will describe the region as
\begin{align}
	\mathcal{T}=\bigcup_i\phi(\mathcal{R}_i)\ .
\end{align}

Each region ${\cal R}_i$ is a positive geometry with connected interior and it's  canonical form is  given explicitly as
\begin{align}\label{OMegaRi}
	\Omega(\mathcal{R}_i)=\prod_{j=1}^{km}\left (\frac{1}{x_j-a_j}- \frac{1}{x_j-b_j}\right) dx_j\ .
\end{align}
The oriented canonical form of $\cal T$ itself, is then by definition simply the sum of such contributions from each region ${\cal R}_i$.

The simplest way to see that $\mathcal{R}_i$ is a positive geometry with canonical form given by~\eqref{OMegaRi} is to use the following  simple change of variables:
\begin{align}
	x_j'=-\frac{x_j-a_j}{x_j-b_j}\,.
	\label{map}
\end{align}
Then the inequalities describing $\mathcal{R}_i$ become simply $x_j'>0 \ \forall j$ which is a positive geometry with canonical form 
\begin{align}\label{simpleCF}
	\Omega(\mathcal{R}_i')= \prod_{j=1}^{km}\frac{d x_j'}{x_j'}.
\end{align}
As discussed the in section 4 of \cite{Arkani-Hamed2017}, if we have a rational map  from a positive geometry $\mathcal{R}'$ to a positive geometry $\mathcal{R}$, then the canonical form of $\mathcal{R}$ is the push-forward of the canonical form of $\mathcal{R}$.  This means that if in \eqref{map} $a_j(x)$ and $b_j(x)$ are rational functions, we can then rewrite the form back in the $x$ variables and obtain \eqref{OMegaRi}. The inequalities describing the amplituhedron-like geometries are such that they are always multi-linear for coordinates for which the entries of $Y$ are multi-linear.

As you can see, besides computing the canonical form of $\mathcal{R}_{i}$ this algorithm also assigns to the canonical form a precise sign.
If we compute the canonical form of two regions that share a codimension 1 boundary we want the residue of the two canonical forms on that boundary to cancel. Let's see how this sign choice automatically fulfills this requirement. Suppose we have two regions $\mathcal{R}_1$ and $\mathcal{R}_2$ described by inequalities of the form \eqref{easyregion} that also share a codimension 1 boundary $\mathcal{B}$. By the definition of the canonical form we know that the residue on  $\mathcal{B}$ of these two forms will be the same up to a sign. This means that if we choose local coordinates $\{z,x_1,\cdots\}$ such that the shared boundary lies at $z=0$, we have 
\begin{align}
	\Omega(\mathcal{R}_{1,2})=\pm \omega(\mathcal{B}) \frac{dz}{z} + O(z^0), 
\end{align}
where $\omega(\mathcal{B})$ is the canonical form of the shared boundary and  $o(z^0)$ is the non divergent part for $z=0$ of the canonical form. 
Now we just need to prove that the residue of the two forms have opposite sign. Observe that each canonical form \eqref{simpleCF}, once we strip the differential, is positive inside its region. The two regions are on two different sides of $\mathcal{B}$, that is $z<0$ and $z>0$. Therefore one form must be positive for $z<0$ and the other must positive for $z>0$, i.e. they have  opposite signs.

To summarize, given a coordinate patch $\Phi:\ \mathbb{R}^{km} \rightarrow \widetilde{Gr}(k,k+m)$ for each region of the form \eqref{easyregion} covered by the patch the CAD algorithm gives a canonical form with a well defined sign. The relative sign of the canonical forms of any adjacent regions, that is two regions that share a codimension 1 boundary, is such that their residues cancel on shared codimension 1 boundaries. 

This implies that the algorithm assigns compatible orientations to adjacent regions. Notice that the fact that all adjacent regions have compatible orientations and the fact that we can triangulate the whole orientable Grassmannian with the CAD implies that all the regions have the same orientation, that is the CAD algorithm gives the globally oriented canonical form.

Note that when applying the CAD algorithm it is important to check the orientation of the coordinate map. If the sign of the measure, $\prod_{i}
\braket{Y d^m Y_i}$ ( see footnote~\ref{f5}) is positive, the coordinate chart is orientation preserving, if negative it is orientation reversing and if zero it is degenerate. The reversed orientation contributes with minus sign to the canonical form. If the measure vanishes anywhere in the geometry, we need to split it into regions where the measure is everywhere non-vanishing and sum the result for the different regions with the sign contributions coming form the sign of the measure. 

Let's consider as an example the computation of the oriented canonical form of the 4-points one loop squared amplituhedron $\sH_{4,0,1}$. Examining the definitions in sections~\ref{amplike} and~\ref{SquaredAmplituhedron}
for  this case we have that $f=0$ and $l'=0$ or 1. The squared amplituhedron is  $\sH_{4,0,1}=\sH^+_{4,0,1}$ (there is no $\sH^-_{4,0,1}$) and 
$\sH^+_{4,0,1}=\sH^{(0,0)}_{4,0,1}\cup \sH^{(0,1)}_{4,0,1}$. 
The second of these $ \sH^{(0,1)}_{4,0,1}=\sA_{4,0,1}$ is the standard amplituhedron geometry and has loop flipping number $f_1=2$ (recall from~\eqref{lprime} that $l'$ is the number of loops that have flipping numbers $f{+}2$ - here we thus have $l'=1$ loop variable with flipping number $f{+}2=2$).
The geometries live in $\widetilde{Gr}(2,4)$ and are defined by the physical inequalities 
\begin{align}
	\braket{AB12}>0, \qquad \braket{AB23}>0,\qquad \braket{AB34}>0,\qquad \braket{AB14}>0
\end{align}
together with inequalities arising from the two different flipping numbers: 
\begin{align}
	\braket{AB13}>0 \text{\ (for\ } \sH^{(0,0)}_{4,0,1})\qquad \qquad     \braket{AB13}<0 \text{\ (for\ } \sH^{(0,1)}_{4,0,1})\ .
\end{align} 
These two geometries are almost disconnected, that is they do not share any codimension 1 boundary. If we fix $Z$ to the identity and choose coordinates for $AB$
\begin{align}
	AB=\begin{pmatrix} 1 & x & 0 &-w\\ 0 & 1 & y & z\end{pmatrix},
\end{align}
we see that the parametrized inequalities describing the amplituhedron, $\sH^{(0,1)}_{4,0,1}$,  are $x,y,w,z>0$ and the one describing $\sH_{4,0,1}^{(0,0)}$ are $x,y,w<0$ and $z>0$. These two regions are positive geometries and the cylindrical decomposition prescription~\eqref{OMegaRi} appears to yield
$\frac{dx}{x}\frac{dy}{y}\frac{dw}{w}\frac{dz}{z}$ for the former and  $-\frac{dx}{x}\frac{dy}{y}\frac{dw}{w}\frac{dz}{z}$ for the latter. However if we look at the measure we see that
\begin{align}
	\braket{AB d^2A}\braket{ABd^2B}=y\, dx\,dy\,dz\,dw,
\end{align}
which changes sign according to whether $y>0$ or $y<0$. These means that the coordinates we chose are orientation reversing for $y<0$.
To obtain the oriented canonical form we therefore have to multiply by a $-1$ the result of the CAD on $\sH_{4,0,1}^{(0,0)}$ and the final result for the global oriented canonical form of $\sH_{4,0,1}$ is therefore
\begin{align}
	H_{4,0,1} = 2\frac{dx}{x}\frac{dy}{y}\frac{dw}{w}\frac{dz}{z}= 2 \frac{\braket{1234}^2\braket{AB d^2A}\braket{ABd^2B}}{\braket{AB12}\braket{AB23}\braket{AB34}\braket{AB14}},
\end{align}
which corresponds to  2 times the 4 point 1 loop MHV amplitude as expected.

Finally we give another simple example which illustrates what happens when we need multiple coordinate charts to cover the space. Consider the triangle  defined by $Z_1=(1,0,1)$, $Z_2=(1,0,-1)$, $Z_1=(0,1,0)$. On the oriented projective space (equivalent to a sphere) the coordinate chart $Y=(x,y,1)$ does not contain this triangle ($Z_2$ clearly lies outside this coordinate chart) so we need the additional chart $Y=(x',y',-1)$. The first chart covers the northern hemi-sphere and the second the southern hemisphere.
The triangle is defined by the inequalities $\braket{Y i\,i{+}1}>0$. In the first chart this gives the region $y>0,x>1$, yielding canonical form $dx dy/((x-1)y)$ and in the second chart it gives the region $y'>0,x'>1$, yielding canonical form $-dx' dy'/((x'-1)y')$. The additional minus sign in the second case arises from the orientation of the coordinates map $\braket{Y d^2 Y}= 2dxdy=-2dx'dy'$, negative in the second chart. 
Now {\em after} we have obtained the canonical forms in the two charts we need to add them together.
One way to do this  is to covariantise the two forms above and then add them together. In order to covariantise it will be useful to  introduce the additional vertex $Z_*=(1,0,0)$, the point where the boundary of the triangle meets the equator hence moving from the northern to the southern hemisphere. The result will just be the sum of the two triangles obtained by splitting the big triangle along the equator (see picture).
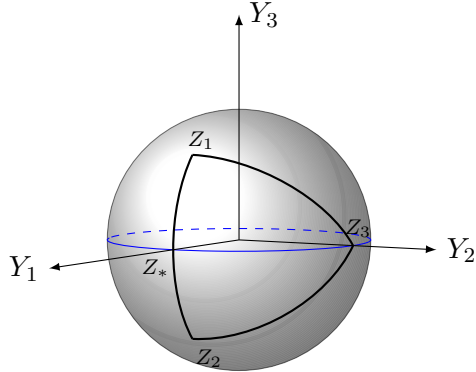
\begin{center}
	\begin{tikzpicture}[scale=1]
		\pgfmathsetmacro\R{sqrt(3)} 
		\draw[ball color=white,opacity=.5] (0,0,0) circle (\R); %
		\tdplotsetmaincoords{85}{120}
		\begin{scope}[tdplot_main_coords, shift={(0,0)}]
			\coordinate (O) at (0,0,0);

			\pgfmathsetmacro{\thetavec}{0}
			\pgfmathsetmacro{\phivec}{0}
			\tdplotsetrotatedcoords{\phivec}{\thetavec}{0}
			\tdplotdrawarc[tdplot_rotated_coords,color=blue]{(O)}{\R}{-70}{110}{}{}
			\tdplotdrawarc[tdplot_rotated_coords,color=blue, dashed]{(O)}{\R}{110}{290}{}{}
			\node[yshift=4 mm] at (1,0,1) {\textcolor{black}{\scriptsize 
					$Z_1$}};
			
			\pgfmathsetmacro{\thetavec}{45};
			\pgfmathsetmacro{\phivec}{0}
			\tdplotsetrotatedcoords{\phivec}{\thetavec}{0};
			\tdplotdrawarc[tdplot_rotated_coords,color=black,thick]{(O)}{\R}{0}{90}{}{};
			\node[yshift=-5 mm,xshift=1mm] at (1,0,-1) {\textcolor{black}{\scriptsize $Z_2$}};
			
			\pgfmathsetmacro{\thetavec}{45};
			\pgfmathsetmacro{\phivec}{180}
			\tdplotsetrotatedcoords{\phivec}{\thetavec}{0};
			\tdplotdrawarc[tdplot_rotated_coords,color=black,thick]{(O)}{\R}{180}{270}{}{};
			\node[yshift=2 mm,xshift=7mm] at (0,1,0) {\textcolor{black}{\scriptsize $Z_3$}};
			\node[yshift=-3 mm,xshift=-6mm] at (1,0,0) {\textcolor{black}{\scriptsize $Z_*$}};
			\pgfmathsetmacro{\thetavec}{90};
			\pgfmathsetmacro{\phivec}{90}
			\tdplotsetrotatedcoords{\phivec}{\thetavec}{0};
			\tdplotdrawarc[tdplot_rotated_coords,color=black,thick]{(O)}{\R}{225}{315}{}{};

			\coordinate (X) at (5,0,0) ;
			\coordinate (Y) at (0,3,0) ;
			\coordinate (Z) at (0,0,3) ;
			
			\draw[-latex] (O) -- (X) node[anchor=east] {$Y_1$};
			\draw[-latex] (O) -- (Y) node[anchor=west] {$Y_2$};
			\draw[-latex] (O) -- (Z) node[anchor=west] {$Y_3$};
			
		\end{scope}
	\end{tikzpicture}
\end{center}

However it is also possible to  add the two forms together directly at the level of coordinates. To do this we first realise that now,  at the level of the form,  we can safely project from the sphere to $P^2$. For the second chart we then have that $(x',y',-1)\sim (-x',-y',1)$ and we can map safely back  to  $(x,y)$ coordinates as $x=-x',y=-y'$. We then have the second form directly in $x,y$ coordinates as 
$-dx' dy'/((x'-1)y')=-dx dy/((1+x)y)$. Now we are using the same co-ordinates for both terms, we can safely sum the two contributions together. Finally we can  covariantise the final result if we like giving
\begin{align}
	\frac{dx\,dy}{(x-1)y}-\frac{dx\,dy}{(x+1)y} =\frac{2dx\,dy}{(x-1)(x+1)y} = \frac{\braket{Y d^2Y}\braket{123}^2}{\braket{Y12}\braket{Y23}\braket{Y31}}\ ,
\end{align}
the correct canonical form for the triangle.

\section{Proof and checks of  the conjectures}
\label{Proof-checks}

In this section we examine the equivalence of the two definitions of amplituhedron-like geometries, find a triangulation of the amplituhedron-like geometry via pairs of on-shell diagrams and  use this to formulate a proof of the main conjecture~\eqref{conjecture1},\eqref{conjecture2} that the amplituhedron-like geometries give products of amplitudes at tree level and also at loop level in the MHV case. Note that all the proofs 
assume the truth of various conjectures regarding the amplituhedron itself.

\subsection{Equivalence of definitions of amplituhedron-like geometries}
\label{AmpAsProd}

In  section~\ref{altdef}  we proposed an alternative definition for the amplituhedron-like geometry  as the image of two positive Grassmannians for $k=n-m$. This definition has the nice feature that it apparently manifests the product structure of amplituhedron-like geometries observed on taking the canonical form \eqref{conjecture1}.   
Here we prove that this alternative definition~\eqref{GrassmannianProduct} is a subset of the sign flip definition~\eqref{amplituhedronlike}
\begin{equation}
	\sH_{n,n-m}^{(f);\text{alt}}   \subseteq \sH_{n,n-m}^{(f)} \ .
\end{equation}	

So in other words  we want to prove that any $Y$ that can be written as
\begin{equation}\label{91}
	Y=\begin{pmatrix}C_1\\C_2\end{pmatrix}\cdot Z\  \text{ where } C_1\in Gr_{\lessgtr}(f,n) \text{ and }  C_2\in \text{alt}(Gr_{>})(n{-}m{-}f,n)
\end{equation}
must necessarily then satisfy the defining inequalities of the sign flip definition of $\sH_{n,n-m}^{(f)}$.
To do this we first 
split the $(n{-}m)$-plane $Y$ into an $f$-plane $Y_1$ and a $(n{-}m{-}f)$-plane $Y_2$
\begin{align}
	&Y_1=C_1\cdot Z &Y_2=C_2\cdot Z\,,
\end{align}
and consider projecting the geometry  onto $Y_1^\perp$.
Thus we define $\braket{*}_{Y_1}:= \braket{Y_1 *}$,  brackets projected onto $Y_1^\perp$.
Now notice that the projected $Z$'s satisfy
\begin{align}\label{braket2}
	\braket{ Z_J}_{Y_1}:=\braket{ Y_1 Z_J} &=  \Delta_{\bar J}(C_1)\,\braket{Z_{\bar J} Z_J} \nonumber \\
	&=\Delta_{\bar J}(C_1)\braket{1 \cdots n}(-1)^{\#_\text{odd}(J) }(-1)^{g_{n,f}} \,,
\end{align}
where  $J$ is an ordered list of $n{-}f$ elements,
$\bar J$ is the ordered complement of $J$ in $1,..,n$,  $\#_\text{odd}(J)$ is the number of odd elements  in $J$ and $g_{n,f}:=	\lfloor\frac{n-f}{2}\rfloor+(n{-}f)n$ introduced in~\eqref{altdef}. The second equality arises simply from reordering $J,\bar J$:  first reverse the order of $J$ (introducing the factor $(-1)^{\lfloor\frac{n-f}{2}\rfloor}$ of $g_{n,f}$) and then permute sequentially the elements of the reversed $J$ starting from the leftmost, into  the correct position to obtain the remaining terms.  Now defining  $\tilde Z_i= (-1)^i Z_i$, since the  ordered minors of $C_1$ are positive or negative according to the sign of $(-1)^{g_{n,f}}$~\eqref{altdef}, we obtain that the projected ordered $\tilde Z$s are totally positive
\begin{align}\label{braket}
	\braket{ \tilde Z_J}_{Y_1}>0\,.
\end{align}
Since $C_2 \in \text{alt}(Gr)_>(n-f,n)$ then $ (\tilde C_2)_{\alpha i} : = (-1)^i (C_2)_{\alpha i} \in Gr_>(n-f,n)$ and we have that $Y_2= C_2.Z= \tilde C_2.\tilde Z$.
Thus $Y_2$ and the $\tilde Z$s, both projected onto $Y_1^\perp$, give a geometry  equivalent to the amplituhedron  $\sA_{n,n-m-f}(Y_2;\tilde Z)$~\eqref{amplituhedron}. 

But then this means the projected $Y_2$  must satisfy the conditions of the equivalent sign flip definition of this amplituhedron.%
\footnote{It is still conjectural that the two definitions of the amplituhedron are equivalent but it has been proven that the original definition is a subset of the sign flip definition~\cite{Arkani-Hamed2018} which is all we need here.} 
So for example  taking $m=4$ for concreteness (but one can check the general case similarly)  the projected brackets satisfy the sign flip definition of $\sA_{n,n-m-f}(\tilde Z)$:
\begin{align}\label{br1}
	\braket{Y_2 \tilde Z_i \tilde Z_{i+1}\tilde Z_j \tilde Z_{j+1}}_{Y_1}>0, \quad (-1)^{n-f}\braket{Y_2 \tilde Z_i \tilde Z_{i+1}\tilde Z_1\tilde Z_n}_{Y_1}>0, \quad \{\braket{Y_2 \tilde Z_1\tilde Z_2\tilde Z_3 \tilde Z_i}_{Y_1}\} \begin{array}{ll}
		\text{ has $n{-}4{-}f$}   &  \\
		\text{ sign flips} & 
	\end{array}\ 
\end{align}
Therefore back in the full geometry, switching to the $Z$s, this becomes
\begin{align}\label{br2}
	\braket{Y  Z_i  Z_{i+1} Z_j  Z_{j+1}}>0, \quad (-1)^{f}\braket{Y  Z_i  Z_{i+1} Z_1 Z_n}>0, \quad \{\braket{Y  Z_1 Z_2 Z_3  Z_i}\} \text{ has $f$ sign flips}\ 
\end{align}
which are just the defining inequalities showing that   $Y \in \sH_{n,n-m}^{(f)}(Z)$.

So we have proved that the alternative definition of amplituhedron-like geometry lies inside the sign flip definition,  $ \sH_{n,n-m}^{(f);\text{alt}}   \subseteq \sH_{n,n-m}^{(f)} $. To show equivalence we also therefore need to show the converse $ \sH_{n,n-m}^{(f)}   \subseteq \sH_{n,n-m}^{(f;\text{alt})} $.
We have been unable to prove this in general (indeed this is similar to the situation for the two equivalent descriptions of the amplituhedron itself where only one direction has been proven) so leave it conjectural.

Note that it is enough to prove that for any $Y \in \sH_{n,n-m}^{(f)}$ there exists a  $C_1\in Gr_{>}$ such that $Y=(C_1, C_2)^T . Z$. It then follows automatically that there exists a $C_2\in \text{alt}(Gr_{>})$ such that $Y_2=C_2.Z$ using essentially the same logic as above.
Indeed
if $C_1$ is positive then $ Y \in \sH_{n,n-m}^{(f)}$ implies $Y_2\in \sA_{n,n-m-f}(\tilde Z)$ (since clearly \eqref{br1} $\Leftrightarrow$ \eqref{br2}). Therefore there exists a $\tilde C_2\in Gr_{>}$ such that $Y_2 =\tilde C_2.\tilde Z$ (here we are assuming that both definitions of the amplituhedron are equivalent) then letting $ ( C_2)_{\alpha i} : = (-1)^i (\tilde C_2)_{\alpha i}$ gives such a $C_2$.
However we have been unable to prove in general that there always exists such a positive $C_1$.

Instead then let us  show this converse statement explicitly in the simplest example of $n=6, k=2,f=1$. Here we initially gauge fix the $C$-matrix  as
\begin{align}\label{ineq}
	C= \left( \begin{array}{cccccc}
		x &1&y&b&0&a  \\
		-c&0&-d&z&-1&w 
	\end{array} \right)\ .
\end{align}
Imposing the physical inequalities (first two lines of~\eqref{amplituhedronlike}) gives the inequalities
\begin{align}\label{98}
	a,b,c,d&>0\notag\\
	yz+bd&>0\notag\\
	xw+ac&>0\notag\\
	xz+bc&>0\notag\\
	yw+ad&>0\ .
\end{align}
We now split the space into three regions and perform the following $SL(2)^+$ transformations in each  region to ensure that $C_1$ (the first row of $C$) is strictly positive) and thus of the form \eqref{GrassmannianProduct}
\begin{align}
	x&>0,y>0 \qquad &C\rightarrow \left( \begin{array}{cc}1 &-\epsilon\\\epsilon'&1\end{array} \right)C = &\left(\begin{array}{cccccc}x &1&y&b&\epsilon&a  \\-c&\epsilon'&-d&z&-1&w\end{array} \right)\notag\\
	x&<0,y>\tfrac{dx}c \qquad &C\rightarrow \left( \begin{array}{cc}
		1 &\tfrac xc{-}\epsilon  \\
		\epsilon'&1
	\end{array} \right)C = &\left( \begin{array}{cccccc}
		c \epsilon &1&y-\tfrac{dx}c&\tfrac{xz+bc}c&\tfrac{-x}c&\tfrac{xw+ac}c  \\
		-c&\epsilon'&-d&z&-1&w 
	\end{array} \right)\notag\\
	y&<0,x>\tfrac{cy}d \qquad &C\rightarrow \left( \begin{array}{cc}
		1 &\tfrac yd{-}\epsilon  \\
		\epsilon'&1
	\end{array} \right)C = &\left( \begin{array}{cccccc}
		x-\tfrac{cy}d &1&\epsilon d&\tfrac{yz+bd}d&\tfrac{-y}d&\tfrac{yw+ad}d  \\
		-c&\epsilon'&-d&z&-1&w 
	\end{array} \right)
\end{align}
Here the variables $\epsilon,\epsilon'$ are positive but small enough so that their presence does not change the sign of any non-zero entries in the $C$-matrix. For simplicity we have omitted such terms in the non-zero entries of the matrix. Their job is simply to move from the boundary of the region to the interior.  
We now observe that in all three cases (using the inequalities \eqref{ineq}) the top row of $C$ is indeed positive. Therefore we know in advance that the second row of $C$ must be in $\text{alt}(Gr)_>(1,6)$ and indeed that is what we find. 
So we have shown in this example  that indeed for any $Y$ in the amplituhedron-like geometry (which gives~\eqref{98}) we can find $C_1,C_2$ such that $Y$ has  the form~\eqref{91}.

\subsection{On-shell diagrams}
\label{On-shell diagrams}

The superamplitude can be computed by summing a certain set of  on-shell diagrams~\cite{Arkani-Hamed:2016byb}. Each on-shell diagram has a geometrical interpretation and the corresponding union of geometries then  yields a triangulation of the corresponding amplituhedron. 
In this section we will make a similar claim for the amplituhedron-like geometries.

First we quickly review the key points we need from the standard on-shell diagram story for amplitudes. 
Each on-shell diagram is completely characterized by an affine (or decorated) permutation $\sigma$, which maps points $a \in 1,..,n$ to $\sigma(a)$ where $a \leq \sigma(a) \leq a+n$. 
Each permutation, in turn, identifies a specific parametrisation of a matrix $C_\sigma(\alpha)$ in the oriented Grassmannian  $\widetilde{Gr}(k,n)$, that is the set of $k\times n$ matrices modulo a $GL_+(k)$ transformation, where $k$ is the number of $a$ such that $\sigma(a)>n$. The evaluation of any on-shell diagram in momentum supertwistor space,  labelled by an affine permutation $\sigma$, can then be written as
\begin{align}\label{osd0}
	f_\sigma^{(k)}= \int \frac{d \alpha_1}{\alpha_1}\cdots  \frac{d \alpha_{4k}}{\alpha_{4k}}\delta^{(4|4)\times k}(C_\sigma(\alpha)\cdot \Z)\,.
\end{align}
Any $C_\sigma(\alpha)$ generated from an affine permutation $\sigma$ has the property that for $\alpha_i>0$ all its minors are $\geq 0$. The space of all elements in $\widetilde{Gr}(k,n)$ with non negative minors is called the non-negative Grassmannian and is denoted Gr$_{\geq0}(k,n)$.  So, for each affine permutation $\sigma$ we can define a region $\Pi_\sigma^>= \{C_\sigma(\alpha):\alpha_i>0\}$ in Gr$_{\geq0}(k,n)$ called a positroid cell.

How is this connected with the amplituhedron and its canonical form?
We know that the amplituhedron can be defined as the image of the positive Grassmannian through a map $Y=C\cdot Z$, where $Z\in$Gr$_{>}(k,k+4)$. Consider a set of on-shell diagrams labelled by the  affine permutations $\sigma_i$ that give the N$^k$MHV amplitude. Then the images of the corresponding positroid cells $Z(\Pi_{\sigma_i})$, that is the regions parametrised by $Y_{\sigma_i}(\alpha)=C_{\sigma_i}(\alpha)\cdot Z$ for $\alpha_i>0$, triangulate the amplituhedron. Moreover the integrand  of the on-shell diagram~\eqref{osd0} is  the canonical form of the image of the positroid cell in the coordinates $Y=C_{\sigma_i}(\alpha)\cdot Z$
\begin{align}
	\Omega(\Pi_{\sigma_i})=\frac{d \alpha_1}{\alpha_1}\cdots  \frac{d \alpha_{4k}}{\alpha_{4k}}\ .
\end{align}
Thus we can compute the amplituhedron canonical form by summing the positroid canonical forms.

Now consider the product of two on shell diagrams $f_{\sigma}^{(k_1)}$ and $f_{\tau}^{(k_2)}$
\begin{align}
	f_{\sigma}^{(k_1)}f_{\tau}^{(k_2)}= \int \frac{d \alpha_1}{\alpha_1}\cdots  \frac{d \alpha_{4k_1}}{\alpha_{4k_1}}\frac{d \beta_1}{\beta_1}\cdots  \frac{d \beta_{4k_2}}{\beta_{4k_2}}\delta^{(4|4\times k)}\left( \begin{psmallmatrix}
		C_{\sigma}(\alpha)\\C_{\tau}(\beta)
	\end{psmallmatrix}\cdot \Z \right).
	\label{onshellProd}
\end{align}
where $k=k_1+k_2$.
This equation makes manifest that the product of two or more on-shell diagrams has only dlog singularities and maximal residues equal to $\pm 1$, implying that the product of amplitudes has also only dlog singularities.

Now we would like to associate a corresponding geometry in the auxiliary Grassmannian $\widetilde Gr(k,n)$ to the product of on-shell diagrams.  The naive choice would be to consider the region parametrised by $\begin{psmallmatrix}
	C_{\sigma}(\alpha)\\C_{\tau}(\beta)
\end{psmallmatrix}$ for $\alpha_i,\beta_i>0$.
On the other hand, for this to lie in the amplituhedron-like geometry, using the alternative definition~\eqref{GrassmannianProduct}, we should rather have $C_{\sigma}$  in the positive Grassmannian,%
\footnote{Or $Gr_<(k_1,n)$ for $g_{n,f}$ odd. 
	When $g_{n,f}$ is odd we will need to eg flip the sign of one row of $C_{\sigma}(\alpha)$ so it will become an element of the negative Grassmannian. However we will surpress this case from now on for simplicity of presentation, but it is to be understood.}
$Gr_>(k_1,n)$ but $C_{\tau}$ in the {\em alternating} Grassmannian $\text{alt}(Gr_>)(k_2,n)$. From this perspective it is thus  natural  to associate to the product of two on-shell diagrams characterized by the auxiliary matrices $C_{\sigma}(\alpha)\in Gr_\geq(k_1,n)$ and $C_{\tau}(\beta)\in Gr_\geq(k_2,n)$,  a region $\Pi_{\sigma,\tau}^>$ defined as
\begin{align}\label{pi}
	\Pi_{\sigma,\tau}^>:= \{C_{\sigma,\tau}=\begin{pmatrix}
		C_{\sigma}(\alpha)\\ \text{alt}(C_{\tau})(\beta)\end{pmatrix} \quad \text{for} \quad \alpha_i,\beta_i>0\}
\end{align}
where $\text{alt}$ flips the sign of the odd columns of $C_2$ (which will not affect~\eqref{onshellProd}).

Note that the product of on-shell diagrams can vanish ( for example the product of identical on-shell diagrams must vanish). In these cases the corresponding geometry $\Pi_{\sigma,\tau}^>$ is not full dimensional.

The canonical form of this geometry, in the coordinates $Y=C_{\sigma,\tau}.Z$, is then the integrand in~\eqref{onshellProd} 
\begin{align}
	\Omega(Z(\Pi_{\sigma,\tau}^>))= \frac{d \alpha_1}{\alpha_1}\cdots \frac{d \alpha_{4k_1}}{\alpha_{4k_1}}\frac{d \beta_1}{\beta_1}\cdots  \frac{d \beta_{4k_2}}{\beta_{4k_2}}\ .
\end{align}
However,  although the corresponding expression in superspace is a standard product still, we therefore know that the canonical form must give the star product of the separate covariantised forms
\begin{align}\label{osdstar}
	\Omega(Z(\Pi_{\sigma,\tau}^>))=  \Omega(Z(\Pi_{\sigma}^>)) * \Omega(Z(\Pi_{\tau}^>))\ .
\end{align}

\subsection{Proof of the conjecture at tree-level}
We now have all the ingredients needed to prove that amplituhedron-like geometries yield products of amplitudes~\eqref{conjecture1}.
Consider two sets of on-shell diagrams $\{f^{(k_1)}_{\sigma_{i}}\},\{f^{(k_2)}_{\tau_j}\}$ which each sum to separate (parity conjugate) amplitudes
\begin{align}\label{osd}
	\mathcal{A}_{n,k_1}=\sum_if^{(k_1)}_{\sigma_{i}},\qquad \mathcal{A}_{n,k_2}=\sum_j f^{(k_2)}_{\tau_j},
\end{align}
with $k=k_1+k_2=n-m$. We would like to prove that the set of all the associated geometries 
$Z(\Pi_{\sigma_i,\tau_j})$ (defined in~\eqref{pi}) is a triangulation of the corresponding amplituhedron-like geometry $\sH_{n,n-m}^{(k_1);\text{alt}}$.  That is we wish to show its elements are disjoint and their union covers the amplituhedron-like geometry. Since the oriented canonical form of a triangulation is given by the sum of the canonical forms of its elements, this then  automatically proves that this geometry yields the product of amplitudes~\eqref{conjecture1}. 

To do this we will prove that for every $Y\in \sH_{n,n-m}^{(k_1);\text{alt}}$, $Y$ belongs to a unique region $Z(\Pi_{\sigma_{i^*},\tau_{j^*}})$ (defined in~\eqref{pi}). That is there exist unique indices $i^*,j^*$ such that $Y$  can be written as $Y=Y_1Y_2$ with $Y_1=(C_{\sigma_{i^*}}(\alpha)) \cdot Z $ and $Y_2 = \text{alt}(C_{\tau_{j^*}}(\beta)) \cdot Z$ for some $\alpha,\beta>0$, where $C_{\sigma_{i^*}}(\alpha),C_{\tau_{j^*}}(\beta)$ are the $C$ matrices associated with the corresponding on-shell diagrams $f_{\sigma_{i^*}}^{(k_1)},f_{\tau_{j^*}}^{(k_2)}$ respectively in~\eqref{osd}.  

So we start with an arbitrary $Y$ in the  amplituhedron-like geometry, $\sH_{n,n-m}^{(f);\text{alt}}$~\eqref{GrassmannianProduct},  so 
\begin{equation}\label{start}
	Y=Y_1Y_2, \qquad \text{with} \quad \begin{pmatrix}Y_1\\Y_2\end{pmatrix}=\begin{pmatrix}C_1\\C_2\end{pmatrix}\cdot Z
\end{equation}
for some $C_1\in Gr_{>}(k_1,n)$ and $C_2\in \text{alt}(Gr_{>})(n{-}m{-}k_1,n)$.
We then follow the first part of the argument in section~\ref{AmpAsProd}. Namely we project onto a $(n{-}k_1)$-plane orthogonal to $Y_1$, $Y_1^\perp$,  and note that the resulting geometry of the projected $Y_2$ is the amplituhedron $\sA_{n,n-m-k_1}((Y_2)_{Y_1},(\tilde{Z})_{Y_1})$,  projected on $Y_1^\perp$ and in terms of alternating ($\tilde Z_i:= (-1)^iZ_i$) and  projected external data $( \tilde Z)_{Y_1}$ (see the paragraph containing~\eqref{braket}). Here the subscript simply denotes the projection on $Y_1^\perp$. We then use the fact that we know that this amplituhedron can be described geometrically as the disjoint union of on-shell diagrams in $Gr(n{-}m{-}k_1,n{-}k_1)$, the space of $(n{-}m{-}k_1)$-planes in the $n{-}k_1$ subspace $Y_1^\perp$. Therefore there exists a unique $j^*$ such that the projection of $Y_2$ on $Y_1^\perp$ can be written as $(Y_2)_{Y_1}=C_{\tau_{j^*}}(\beta)\cdot ( \tilde Z)_{Y_1}$ for some $\beta>0$. Now comes the key part of the proof: we can then project  back away  from the hyperplane $Y_1^\perp$ 
by defining  $\widehat{Y}_2=C_{\tau_{j^*}}(\beta)\cdot \tilde Z =\text{alt}(C_{\tau_{j^*}})(\beta)\cdot Z$. (In the second equality we have simply swapped the flipping of odd particles from the $\tilde Z$ to the $C$ matrix). We have now that $Y=Y_1Y_2=Y_1\widehat{Y}_2$.

Now we can do a similar manipulation,  but now projecting the geometry (both $Y_1$ and the $Z$s) onto the $(n{-}k_2)$-plane $\widehat{Y}_2^\perp$.  Following similar logic to that of~\eqref{braket2} we find that $(Y_1)_{\widehat{Y}_2}$ must live in the amplituhedron $\sA_{n,k_1}((Y_1)_{\widehat{Y}_2},(Z)_{\widehat{Y}_2})$ on $\widehat{Y}_2^\perp$, where $(Z)_{\widehat{Y}_2}$ lives in non negative Grassmannian, that is all its minors are either positive or zero
\footnote{A small but important subtlety appears here in that this amplituhedron on $\widehat{Y}_2^\perp$ may be degenerate in the sense that some of the projected $Z$ brackets $\braket{Z}_{\widehat{Y}_2}\geq 0$ may vanish. Nevertheless the  statement which follows in the main text is still true, the consequence of the degeneracy is simply that some on shell diagrams may vanish and the corresponding geometries be non maximally dimensional.}.
Therefore there exists a unique $i^*$ such that $({Y}_1)_{\widehat{Y}_2}=C_{\sigma_{i^*}}(\alpha)\cdot (Z)_{\widehat{Y}_2}$. This  can then be projected back yielding  $\widehat{Y}_1=C_{\sigma_{i^*}}(\alpha)\cdot Z$ with  $Y=\widehat{Y}_1\widehat{Y}_2$. 

We conclude that any $Y$ satisfying \eqref{start}  belongs to one and only one region associated to the product of on-shell diagrams, one in each of the sums in~\eqref{osd}. 
Therefore the regions $Z(\Pi^>_{\sigma_{i^*}\tau_{j^*}})$  are disjoint and cover the corresponding amplituhedron-like geometry
\begin{align}
	\sH_{n,n-m}^{(k_1)} =  \bigcup_{i,j} Z(\Pi^>_{\sigma_i,\tau_j}).
\end{align}
Finally, putting this together~\eqref{osdstar} we obtain the anticipated result. The oriented canonical form of $\sH_{n,n-m}^{(k_1)}$ is given by the product of amplitudes
\begin{align}
	H_{n,n-m}^{(k_1)} =  \sum_{i,j}    {\Omega}(Z(\Pi^>_{\sigma_i,\tau_j}))=\sum_{i}    {\Omega}(Z(\Pi^>_{\sigma_i})*\sum_j{\Omega}(Z(\Pi^>_{\tau_j}))=A_{n,k_1}*A_{n,k_2} \ ,
\end{align}
which concludes our proof.

\subsection{Proof of the loop level conjecture for $f$ maximal}

We can also explicitly prove the loop level conjecture \eqref{conjecture2a} for maximal $f$ flipping number. That is the loop level amplituhedron-like geometry with maximal flipping number 
gives the product of MHV and anti-MHV superamplitudes at all loops,
\begin{align}\label{kk1}
	H_{n,n{-}4,l}^{(n-4;
		\overbrace{\scriptstyle n-2,..,n-2}^{l'},\overbrace{\scriptstyle n-4,..,n-4}^{l-l'}
		)}= A_{n,n-4,l'} \,A_{n,0,l-l'}
\end{align}
The first factor on the RHS, $A_{n,n-4,l'}$, is the anti-MHV $l'$-loop integrand, which itself  factorizes as the tree-level anti-MHV amplitude, $A_{n,n-4,0}$, multiplied by the conjugate of the MHV amplitude $\overline{A_{n,0,l'}}$. Thus we wish to prove 
\begin{align}
	H_{n,n{-}4,l}^{(n-4;
		\overbrace{\scriptstyle n-2,..,n-2}^{l'},\overbrace{\scriptstyle n-4,..,n-4}^{l-l'}
		)}=A_{n,n-4,0}\,\overline{A_{n,0,l'}}\,A_{n,0,l-l'}\ .
	\label{kl}
\end{align}

Nicely  this factorisation can be seen straightforwardly at a purely geometric level. 
Firstly, we can see that the LHS, the loop level anti-MHV amplituhedron-like geometry, is the product of the tree-level anti-MHV amplituhedron, $\sA_{n,n-4}$,  (which $Y$ lies in) and a second  geometry for the loop variables lying in $Y^\perp$, a 4-plane nowhere intersecting any $Y$ in $\sA_{n,n-4}$.
This second geometry turns out to be isomorphic to the $l$-loop MHV amplituhedron-like geometry, with $l-l'$ loops having maximum flipping number $2$ and $l'$ loops having minimum flipping number $0$. Concretely then we first have the geometric factorisation
\begin{align}\label{kl2}
	\sH_{n,n{-}4,l}^{(n-4;
		\overbrace{\scriptstyle n-2,..,n-2}^{l'},\overbrace{\scriptstyle n-4,..,n-4}^{l-l'}
		)}\big(Y,(AB)_i;Z\big)= \sA_{n,n-4}\big(Y;Z\big) \times
	\sH_{n,0,l}^{(0;
		\overbrace{\scriptstyle 0,..,0}^{l'},\overbrace{\scriptstyle 2,..,2}^{l-l'}
		)}
	\big(-(AB)_{i};\tilde Z\big)\ ,
\end{align}
where $\tilde Z_i=(-1)^iZ_i$.  This factorisation can be seen straightforwardly by simply examining the explicit definitions of the geometries involved~\eqref{amplituhedronlikeloop}.
Indeed  $Y$ must lie in the tree  anti-MHV amplituhedron, $Y\in \sA_{n,n-4}$, this is just the first line of the definition of the loop amplituhedron~\eqref{amplituhedronlikeloop}. Then the 2-planes $(AB)_i$ naturally live on the 4-plane, $Y^\perp$, with effective 4-brackets defined as $\braket{*}_Y:=\braket{Y*}$. The resulting effective 4-brackets involving $Z$s then have maximal flipping number $n-4$. Crucially the resulting inequalities are enough to fix all effective $Z$ 4-brackets to be alternating positive:
\begin{align}
	(-1)^{i+j+k+l}\braket{ijkl}_Y>0\qquad \qquad 1{\leq} i{<}j{<}k{<}l{\leq} n\,
\end{align} 
or equivalently $\tilde Z_j:=(-1)^j Z_j$ has positive ordered effective brackets.%
\footnote{Note that one might wonder why a  simple factorisation of geometries like~\eqref{kl2} does not occur for more general amplituhedron-like geometries (ie for lower values of the flipping number $f$). This is because in general there is no simple map $Z \to \tilde Z$s such that all the effective $\tilde Z$-brackets are positive as there is here.} Finally, examining the inequalities $\braket{Y(AB)_i j j{+}1} = \braket{-(AB)_i \tilde j \tilde {k}}_Y(-1)^{j+k+1}$ 
one can check that minimal loop flipping number $n-4$ becomes maximal loop flipping number $2$ and vice-versa. 

Now a second geometrical factorisation occurs  for the second  amplituhedron-like geometry itself,  namely
\begin{align}
	\sH_{n,0,l}^{(0;
		\overbrace{\scriptstyle 0,..,0}^{l'},\overbrace{\scriptstyle 2,..,2}^{l-l'}
		)}=\sH_{n,0,l'}^{(0,0)}\, \sA_{n,0,l-l'}\ .
	\label{loopfact}
\end{align}
Examining the defining inequalities~\eqref{amplituhedronlikeloop}, only the mutual positivity $\braket{AB_iAB_j}>0$ between loops with different flipping number prevents a completely factorised geometry. But a loop $(AB)_j$ with maximal flipping number 2, satisfies the same inequalities as the one loop MHV amplituhedron, and so we can use the original definition of the amplituhedron as the image of the positive Grassmannian to parametrise $(AB)_j$ as
\begin{align}
	(AB)_j= \sum_{l<m} c_{lm} Z_lZ_m,
\end{align}
where $c_{lm}>0$ for $l<m$. Then using this expression we can expand the mutual positivity condition as
\begin{align}
	\braket{(AB)_i(AB)_j}= \sum_{l<m} c_{lm} \braket{(AB)_ilm} \ .
	\label{exp}
\end{align}
Now if $(AB)_i$ has flipping number equal to zero,  all $\braket{(AB)_ilm}$ are positive, implying the positivity of $\braket{(AB)_i(AB)_j}$. As a consequence the geometry  factorizes into the product of $l'$ loops with $f_{AB}=2$ and $l-l'$ loops with $f_{AB}=0$ implying \eqref{loopfact}.

Putting~\eqref{kl2} and \eqref{loopfact} together we arrive at the geometrical double factorisation 
\begin{align}
	\sH_{n,n{-}4,l}^{(n-4;
		\overbrace{\scriptstyle n-2,..,n-2}^{l'},\overbrace{\scriptstyle n-4,..,n-4}^{l-l'}
		)} = 
	\sA_{n,n-4}\, \sA_{n,0,l-l'}\, \sH_{n,0,l'}^{(0,0)}\ ,
\end{align}
which implies (using standard the amplituhedron conjecture together with the fact that the canonical form of geometrical products gives the product of the respective  canonical forms~\cite{Arkani-Hamed2017})
\begin{align}\label{factor1}
	H_{n,n{-}4,l}^{(n-4;
		\overbrace{\scriptstyle n-2,..,n-2}^{l'},\overbrace{\scriptstyle n-4,..,n-4}^{l-l'}
		)} = 
	A_{n,n-4}\, A_{n,0,l-l'}\, H_{n,0,l'}^{(0,0)}\  .
\end{align}
Finally to prove~\eqref{kl} we just need to show that $H_{n,0,l'}^{(0,0)}=\overline{A_{n,0,l'}}$, in other words that the MHV loop amplituhedron-like geometry with all loop flipping numbers {\em minimal} gives  the conjugate of the MHV amplitude. This fact follows nicely from considering the case $l'=l$.   In this case the RHS of~\eqref{factor1} becomes the anti-MHV loop level amplituhedron whose canonical form,  the anti-MHV loop level amplitude, factorises as discussed above~\eqref{kl}. Thus~\eqref{factor1} with $l'=l$ reads $H_{n,n{-}4,l}^{(n-4,l)}=A_{n,n-4,l}=A_{n,n-4} \overline{A_{n,0,l}}   = A_{n,n-4} H_{n,0,l}^{(0,0)}$ and so indeed we have shown that (as conjectured in~\cite{Arkani-Hamed2018} for $l=1$)
\begin{align}\label{amhv}
	H_{n,0,l}^{(0,0)}=\overline{A_{n,0,l}}\ .
\end{align}
This then proves that amplituhedron-like geometries give products of amplitudes at loop level for maximal $k$ and $f$~\eqref{kk1}. 

Note that as a consequence of this derivation we have then proven an interpretation for a particular sector of {\em non-maximal} amplituhedron-like geometries conjectured in~\cite{Arkani-Hamed2018}. Namely the MHV ($k=0$) amplituhedron-like geometries with  arbitrary $n$, from~\eqref{loopfact} and~\eqref{amhv} are given  by
\begin{align}
	H_{n,0,l}^{(0;
		\overbrace{\scriptstyle 0,..,0}^{l'},\overbrace{\scriptstyle 2,..,2}^{l-l'}
		)}=\overline{A}_{n,0,l'}\, A_{n,0,l-l'}\ .
\end{align}

Taking the union over all loop winding numbers to obtain the squared amplituhedron, $\sH_{n,0,l}$, we would expect it to give the sum of these
\begin{align}\label{sqrr}
	H_{n,0,l}= \sum_{l'} \overline{A}_{n,0,l'}\, A_{n,0,l-l'}\ .
\end{align}
Crucially all the almost disjoint amplituhedron-like geometries appearing in the union inherit consistent orientations on the oriented Grassmannian such that they indeed appear with the same sign when taking the globally oriented canonical form and this gives the above  result which is consistent with the square of the amplitude~\eqref{loopsa2}.  
In \cite{Arkani-Hamed2018} it was observed for $n=5$ and conjectured to hold for all $n$ that at one loop 
this union of winding geometries has a (standard) canonical form corresponding to the {\em difference} $A_{n,0,1}-\overline{A_{n,0,1}}$ rather than the sum in~\eqref{sqrr}. This therefore illustrates the importance of the oriented canonical form.

\subsection{Checks of the tree-level general $m$ conjecture}
\label{checksm}
Explicit checks of the amplituhedron-like conjecture can and have been made for various low values of $n,k,l$ on a computer using cylindrical decomposition (see for example~\cite{Eden2017}) but they quickly become too complicated. However the existence of the generalised amplituhedron-like geometries nicely gives another direction in which to to perform checks.

\subsubsection*{Explicit checks for specific values of $m,k,n$}

We have checked the generalised $m$ conjecture~\eqref{conjecture1gen} for $k=2, n=m+2$ and $f=1$ for $m=2,4,6,8$, explicitly, that is 
\begin{align}
	H^{(1)}_{m+2,2}=A_{m+2,1}*A_{m+2,1}.
\end{align}
To do this we first noted that,  $A_{n,1}$, is a natural generalisation of the NMHV amplitude for $m$ even, namely
\begin{align}
	A_{n,1}= \sum_{i_1, \cdots, i_{m/2}} R[1 i_1 i_1{+}1\cdots i_{m/2}i_{m/2}+1]\,,
\end{align}
where
\begin{align}
	R[i_1\cdots i_{m+1}]= \frac{\braket{i_1\cdots i_{m+1}}^m\braket{Yd^mY}}{\braket{Yi_1\cdots i_m}\cdots\braket{Yi_m \cdots i_{m-1}}}
\end{align}
is a generalised $R$-invariant.
We then used this with the formula for the *-product, \eqref{prodRule}, to compute  $A_{m+2,1}*A_{m+2,1}$ covariantly. On the other hand we used the CAD to compute the oriented canonical form of $\sH_{m+2,2}^{(1)}$ and verified that they match.

\subsubsection*{Checks for $m=2$}

For the case $m=2$ the computational complexity is much lower and we have verified \eqref{conjecture1gen} up to $n-2=k=7$. The canonical form for $k=n-2$ reads
\begin{align}
	A_{n,n-2}= \frac{\braket{1,2,\cdots,n}^2}{\prod_{i=1}^{n}\braket{Yii+1}}\ .
	\label{cf_m2}
\end{align}
In \cite{Arkani-Hamed2018} it was proven that for $m=2$, the N${}^k$MHV superamplitude is proportional to the product of $k$  NMHV superamplitudes
\begin{align}
	\mathcal{A}_{n,k}= \frac{(\mathcal{A}_{n,1})^k}{k!}
	\label{allkm2}
\end{align}
and the analogous statement holds in amplituhedron space, so for example for $k=2,m=2$ one can verify that
\begin{align}
	A_{4,2}=\frac{1}{2!}  A_{4,1}*  A_{4,1}=  \frac{\braket{1234}^2}{\braket{Y12}\braket{Y23}\braket{Y34}\braket{Y14}}.
\end{align}
Thus the product of two $m=2$ superamplitudes is
\begin{align}\label{129}
	A_{n,k-k'}*A_{n,k'}= \frac{(A_{n,1})^{*k}}{(k-k')!k'!}= \frac{k!}{(k-k')!k'!}A_{n,k}\ .
\end{align}
We have observed from explicit computations that in fact the geometry in the maximal case, $k=n-2,m=2$, with {\em any} valid sign flip {\em pattern} (ie any specific valid choice of signs for $\braket{Y 123i}$), has a canonical form equal to \eqref{cf_m2}. 
Since for $m=2$, each flipping number $f$ corresponds to  $\begin{psmallmatrix} n{-}2 \\ f \end{psmallmatrix}$ possible flipping patterns (in $n{-2}$ places  you either flip ($f$ times) or don't flip ($n-2-f$ times))
we obtain trivially that
\begin{align}
	H^{(k')}_{n,n-2}=\begin{pmatrix} n{-}2 \\ k' \end{pmatrix} A_{n,k}= A_{n,n-2-k'}*A_{n,k'},
\end{align}
in agreement with the (generalised) amplituhedron-like conjecture \eqref{conjecture1gen}.

\section{Factorisation of sign flip patterns }
\label{factorisation-patterns}

In this section we note a refinement of the factorisation of amplituhedron-like geometries, noting that individual flipping pattern geometries also factorise.

The amplituhedron-like geometries can be divided into regions labelled by a specific sign-flip pattern, that is regions where all brackets $\braket{1,\dots, m-1, i}$ have a well-defined sign. We will indicate the canonical form of a region in $\widetilde{Gr}(k,k+m)$ labelled by a  sign flip pattern ${\bf p}=\{p_1,..,p_f\}$ 
as
$\sh_{n,k}^{\bf p}$. Here
$p_i$ 
denotes  the position of each consecutive  sign flip, so $\text{sgn}(\braket{1,..,m{-}1,p_i-1}=-\text{sgn}(\braket{1,..,m{-}1,p_i}$.
In this notation the canonical form of an amplituhedron-like geometry $\sH_{n,k}^{(f)}$ can be written as
\begin{align}
	\sH_{n,k}^{(f)}=\bigcup_{\substack{\bf p \text{ with}\\  f \text{ flips}}}\sh_{n,k}^{\bf p}.
\end{align}
We have observed by explicit computation for  $m=2,4$ and maximal $k=2=n-m$, that the canonical form of a particular  sign flip pattern geometry $h$ factorises into the following star product, mimicking the same geometrical factorisation mentioned in \eqref{br1}, we have
\begin{align}
	h_{n,n-m}^{\bf p}= h_{n,f}^{\bf p} * A_{n,n-m-f}
	\label{signProd1}
\end{align}
where $f$ is the number of sign flips in the pattern $\bf p$. 
Note that by taking the union over all patterns with a given flipping number, this then implies, and is therefore a refinement of, the main conjecture about amplituhedron-like geometries~\eqref{conjecture1} 
\begin{align}
	H_{n,n-m}^{(f)}= H_{n,f}^{(f)} * A_{n,n-m-f}=A_{n,f} * A_{n,n-m-f}\ .
\end{align}
Now geometries with   complementary sign flip patterns are equivalent, yielding the same canonical form. Indeed clearly the duality relation~\eqref{dualgeometry},\eqref{dualgeometry2} applies to the individual complementary flip pattern geometries, that is
\begin{align}
	h_{n,n-m}^{\bf p}=	h_{n,n-m}^{\bar {\bf  p}}\,,
\end{align}
where $\bar {\bf p}$ indicates the sign flip pattern complementary to $\bf p$. We thus  also have an alternative product formula for a flip pattern geometry 
\begin{align}
	h_{n,n-m}^{\bf p}= h_{n,n-m-f}^{\bar{\bf p}} * A_{n,f}\ ,
	\label{signProd2}
\end{align}
which also implies the main conjecture~\eqref{conjecture1} but this time keeping the other term in the product $A_{n,f}$ whole and reconstructing $A_{n,n-m-f}$.
So one can ``break apart'' either of the two amplitudes appearing in the product but not both simultaneously. It is interesting to compare this with the analogous on-shell diagram story where you can indeed break apart both amplitudes.

We have also observed in all the cases that we have considered that given two flipping patterns ${\bf p}_1$ and ${\bf p}_2$ with flipping number $k'$, the following identity holds
\begin{align} \label{switch}
	h_{n,k'}^{{\bf p}_1} *	h_{n,k-k'}^{\bar{{\bf p}}_2}=		h_{n,k-k'}^{\bar{\bf p}_1}	* h_{n,k'}^{{\bf p}_2} \ .
\end{align}
This relation then implies the equivalence between \eqref{signProd1} and\eqref{signProd2}.

For $m=2$ we can actually prove all these relations from the observation that in the maximal case all flipping patterns yield the same canonical form,  $h^{\bf p}_{n,n-2}=A_{n,n-2}$, as we discussed in section \ref{checksm}. In fact in the non-maximal case we have observed that each individual flipping pattern contributing to the amplituhedron factorises into $k=1$ patterns as follows 
\begin{align}\label{expm2}
	h^{\{p_1,\cdots,p_k\}}_{n,k}&= h_{n,1}^{\{p_1\}}*\cdots* h_{n,1}^{\{p_{k}\}}= \notag \\
	&=\frac{\braket{(1p_1p_1{+}1)(Y\cap1p_2p_2{+}1)\cdots\cap (Y\cap1p_kp_k{+}1)}^2}{\prod_{\alpha=1}^k\braket{Y1p_\alpha}\braket{Y p_\alpha (p_\alpha{+}1)}\braket{Y(p_\alpha{+}1) 1}},
\end{align}
where we used \eqref{multiprodRule} to compute the * product and
\begin{align}
	h_{n,1}^{\{p_i\}}= \frac{\braket{1p_ip_i+1}}{\braket{1p_i}\braket{p_ip_i+1}\braket{(p_i+1)1}}.
\end{align}
For example for the $k=2$ amplituhedron, which has only one sign flip pattern $\{+,-,+\}$,we have
\begin{align}
	A_{4,2}=h_{4,2}^{\{2,3\}}=h_{4,1}^{\{2\}}*h_{4,1}^{\{3\}}=\frac{\braket{123}^2}{\braket{12}\braket{23}\braket{31}}*\frac{\braket{134}^2}{\braket{13}\braket{34}\braket{41}}
\end{align}
Formulas analogous to \eqref{expm2} appear in \cite{Arkani-Hamed:2017mur} and in \cite{Lukowski:2019sxw} and  can be obtained for \eqref{expm2} by expanding $Y\cap (ijk)$ as
\begin{align}
	Y\cap (ijk) = \sum_{\alpha=1}^k (-1)^{\alpha} Y_\alpha \braket{Y_1\cdots Y_{\alpha-1}Y_{\alpha+1}\cdots Y_k ijk}.
\end{align}
If we instead express $Y\cap (ijk)$ as a point on the $3-$plane $ijk$ instead, that is
\begin{align}
	Y\cap (ijk)=\braket{Yij} Z_k-\braket{Yik}Z_j+\braket{Yjk}Z_i,
\end{align}
we obtain an expression for $  \Omega( h^{\{p_1,\cdots,p_k\}}_{n,k})$ where only manifestly $SL(2)$ invariant brackets, that is brackets of the form $\braket{Yij}$, appear.
Note that expression \eqref{expm2} makes  \eqref{switch} trivial for $m=2$.

Now consider the RHS of 
of \eqref{signProd1} and expand the second term into flipping patterns
\begin{align}
	h^{{\bf p}}_{n,f}*A_{n,n-f-2}= h^{{\bf p}}_{n,f}*\sum_{\substack{\bf q \text{ with}\\  f \text{ flips}}} h_{n,n-f-2}^{\bar {\bf q}}.
\end{align}
Now inserting the factorisation~\eqref{expm2}, since $(h_{n,1}^{\{i\}})^{*2}=0$, only the  term ${\bf q}={\bf p}$ in the sum will survive so that there are no repeated factors. Indeed this surviving term will involve  a product over all $n-2$ available flip positions and   we obtain
\begin{align}
	h^{{\bf p}}_{n,f}*A_{n,n-f-2}= h^{\{2\}}_{n,1}*\cdots* h^{\{n-1\}}_{n,1}=A_{n,n-2},
\end{align}
proving the refinement~\eqref{signProd1} of the main conjecture \eqref{conjecture1}  for $m=2$.

\section{Canonicalizing Cyclicity and Crossing}
\label{CanonicalizingCyclicity}

We have seen that the product of two parity conjugate superamplitudes is the canonical form of an amplituhedron-like geometry. 
One could wonder if there are more general geometries which could yield some physical object such as products of two, or more, amplitudes. In particular one could imagine tweaking the signs of the inequalities defining known geometries. At first sight this seems to give  a huge choice of possibilities to investigate.
An obvious property we might insist on to restrict this though is cyclic invariance. In this section we therefore consider the implications of requiring a cyclic invariant canonical form for the corresponding geometry. 
We saw that the amplituhedron-like geometries with even flipping number, $f$, are not cyclic but rather twisted cyclic, $Z_n\rightarrow -Z_1$.  Nevertheless the corresponding canonical form is cyclic, simply due to the fact that  the canonical form is invariant under $Z_1\rightarrow -Z_1$.
It is therefore natural to consider geometries which are cyclic up to any possible flip of the $Z$s. However  in this section we conclude that all such generalised cyclic geometries are equivalent to cyclic or twisted cyclic geometries.
Thus one can define new generalised  geometries by defining arbitrary signs for $\braket{Y12 ii{+}1}$ for each $i$ with cyclicity giving all other physical inequalities from these. On the other hand for the correlator there is the more powerful permutation symmetry and in this case we find a unique correlahedron-like geometry.

\subsection{Cyclic  geometries}

Recall that, as discussed below \eqref{amplituhedron}, it's extremely useful to  consider the geometry 
$Y$ as an {\em oriented} $k$-plane and the $Z$'s as elements in {\em oriented} projective space $ \mathbb{R}^{4k}/GL^+(1)\sim S^{4k-1}$. 
Then we wish to consider geometries $\mathcal{R}(Y;Z_i)$, defined as the set of $Y \in \widetilde{Gr}(k,k+m)$ satisfying a  set of inequalities involving $Y$ and $Z_i$. 
The inequalities will be  invariant under positive rescaling of $Y$ and $Z$ and will be of the form $\braket{Y Z_{i_1}..Z_{i_m}}\lessgtr 0$. Because the canonical form is a rational function, the invariance under positive rescaling of the geometry implies invariance under general rescaling of the canonical form, regardless of the sign of the scaling parameter, i.e. it will be projectively well defined.

This means two very different regions can trivially have the same canonical form: flipping the sign of $Y$ or any  $Z$s, the inequalities defining the geometry will change, while its canonical form will remain the same. 
We thus say that two geometries  $\cR_1,\cR_2$ which are related via such sign flips  are {\em equivalent}, $\cR_1\sim \cR_2$, (and thus
have the same canonical form). 
To this end we would first like to see if all signed cyclic symmetric geometries are equivalent  to cyclic geometries and if not how many inequivalent types of flipped cyclic geometries there are.

Define $F_i$ to be the transformation which flips  $Z_i$,  $Z_i \rightarrow -Z_i$  and 
$F_{I}:=F_{i_1}F_{i_2}...$, where $I:=\{i_1,i_2,\cdots\}$ the transformation that flips the sign of all $Z$'s with index $i\in I$.  Then in this notation the statement of equivalent geometries is that 
\begin{align}
	\cR' \sim \cR \quad \Leftrightarrow \quad  \cR' = F_I \cR \quad \text{for some $I$}\ . 
\end{align}
These transformations clearly satisfy
\begin{align}
	F_I F_I=\mathbb{1},\qquad  
	\mathcal{C}F_I=F_{\mathcal{C}(I)}\mathcal{C},
\end{align}
where $\mathcal{C}$ represents a cyclic transformation, $Z_i \rightarrow Z_{i+1}$.
Now suppose we have a  geometry $\cR$ which is invariant under some flipped cyclicity  $\mathcal{C}F_I$, so $\mathcal{C}F_I \cR=\cR$. A familiar example of this is the twisted cyclicity of the amplituhedron, $Z_i \rightarrow Z_{i+1}$ for $i=1,..,n{-}1$, $Z_n\rightarrow -Z_1$ for which $I=\{n\}$, but we here imagine any possible flipped cyclic geometry.

If we now apply a further $Z$-flip transformation $F_J$ on our geometry $\cR$, then using the above identities  we obtain
\begin{align}
	\mathcal{C}F_I\cR= \cR, \quad &\Rightarrow \quad 
	F_J\mathcal{C}F_I\cR= F_J\cR, \nonumber \\
	&\Rightarrow \quad \mathcal{C}F_{\mathcal{C}^{-1}(J)}F_IF_JF_J\cR= F_J\cR,\nonumber \\
	&\Rightarrow \quad  \mathcal{C}F_{\mathcal{C}^{-1}(J)}F_IF_J\cR'= \cR'
\end{align}
so the equivalent geometry $\cR':=F_J\cR $ is invariant under the flipped cyclicity  $\mathcal{C}F_{\mathcal{C}^{-1}(J)}F_IF_J$. 

A natural question then is whether for any list $I$, we can find a list $J$ such that $F_{\mathcal{C}^{-1}(J)}F_IF_J$ is the identity, and thus obtain an equivalent geometry $\cR'$ which is cyclic invariant (with no flips). Equivalently (by commutativity of $F$ and using $F^2=1$) we ask whether for all $I$ there exists a $J$ such that $F_{\mathcal{C}^{-1}(J)}F_J=F_I$.  
Now if $J=\{j_1,j_2,...\}$ we have 
\begin{align}
	F_{\mathcal{C}^{-1}(J)}F_{J}= (F_{j_1-1}F_{j_2-1}\cdots) (F_{j_1}F_{j_2}\cdots)=(F_{j_1-1}F_{j_1})(F_{j_2-1}F_{j_2}) \cdots 
\end{align}
since the flip operations all commute with each other.
We conclude that $ F_{\mathcal{C}^{-1}(J)}F_{J}$ can be any sign flip transformation with an even number of flips (since any such  can always be constructed from sequences of adjacent flips).
Thus if  $I$ contains an even number of indices, we can always find a list of indices $J$ such that $F_{\mathcal{C}^{-1}(J)}F_I F_J=\mathbb{1}$ and so 
$\cR'=F_J \cR$ is cyclic invariant. 

Instead, if the length of $I$ is odd, but $n$, the total number of indices, is also odd, then  the complementary set $\bar{I}$ will contain an even number of elements. Therefore we can always choose $J$ such that $ F_{\mathcal{C}^{-1}(J)}F_{J}=F_{\bar I}$ and so $ F_{\mathcal{C}^{-1}(J)}F_IF_{J}=F_I F_{\bar I} =F_{\{1,\cdots, n\}}$, so that $\mathcal{C}  F_{\{1,\cdots, n\}}\cR'=  \cR'$.
The transformation $F_{\{1,\cdots, n\}}$ is simply the flipping of all $Z$s and thus for $m$ even will leave all the defining inequalities  $\braket{Yi_1\cdots i_m}\lessgtr 0$ untouched and so we have defined an equivalent cyclic geometry $\cR'$ 
\begin{align}
	\mathcal{C}  F_{\{1,\cdots, n\}}\cR'=\mathcal{C}\cR'=  \cR'.
\end{align}
For $m$ odd we will also have to  flip  the sign of $Y$. 

If the length of $I $ is odd and $n$ is even on the other hand, the best we can do is to chose a $J$ such that only one element is flipped. This is what is known in the literature as twisted cyclicity and one conventionally chooses the element that must be flipped to be $n$ so  $Z_i\to Z_{i+1}$, but $Z_n$ goes to $-Z_1$.

Summarizing the result of our analysis, we can say that, when $n$ is odd (and $m$ even), we can always map any geometry to the cyclic invariant one. When $n$ is even instead we have two classes of geometries: cyclic and twisted cyclic.

Finally we then ask if there is a flip transformations $F_J$ mapping two geometries $\cR_1,\cR_2$  with the {\em same} type of cyclicity $\cC F_I$. Thus we have $\cC F_I\cR_1=\cR_1$, $\cC F_I \cR_2=\cR_2$ and $\cR_2=F_J\cR_1$.
This implies 
\begin{align}
	[F_J,\mathcal{C}]=F_J\mathcal{C}-F_{\mathcal{C}(J)}\mathcal{C}=0 \quad \Rightarrow\quad F_{\mathcal{C}(J)}=F_J.
\end{align}
For a faithful representation of $F_J$ we have just one non-trivial solution,  $F_J=F_{\{1,\cdots, n\}}$. However if the representation of $F_{\{1,\cdots,n\}}=\mathbb{1}$, as is the case of $m$ odd, then we have two further elements in the algebra that commute with $\mathcal{C}$, $ F_J=F_{\text{odd}}=F_{\{1,3,5\cdots\}}$ and $F_J=F_{\text{even}}=F_{\{2,4,6\cdots\}}$. In this representation they correspond to the same operator
\begin{align}
	F_{\text{odd}}F_{\text{even}}=F_{1,\cdots, n} \equiv \mathbb{1} \quad \Rightarrow \quad F_{\text{odd}}=F^{-1}_{\text{odd}}=F_{\text{even}}.
\end{align} 
(More generally we have that $F_I=F_{\bar I}$, where $\bar s$ is the complement of $s$.)
This equivalence of geometries related by $F_{\text{odd}}$ or $F_{\text{even}}$ yields the duality of amplituhedron-like geometries~\eqref{dualgeometry}.

\subsection{Crossing symmetric correlahedron geometries}

In planar $N=4$ SYM there is a class of fundamental  observables that share many properties with amplitudes and have therefore the chance to be defined geometrically. These are the stress energy correlators. These observables can be defined on the twistor on-shell superspace \cite{Chicherin:2014uca}. A point in space time is identified by a line in twistor space, that is a pair of twistors $X_i^{IJ}=Z^1_{iL}Z^2_{iM} \epsilon^{LMIJ}$. In the same way, a point in the chiral super-Minkowski space is identified by a pair of super-twistors.
The supercorrelator can be then organized as a sum over terms with homogeneous Grassmannian degree, usually indicated as $\mathcal{G}_{n,k}$, where $n$ is the number of super twistors and $4(k+n)$ is the Grassmannian degree. In \cite{Eden2017} the  chiral super-Minkowski space is bosonised and the functions $\mathcal{G}_{n,k}$ uplifted to differential forms on the Grassmannian Gr$(k{+}n,k{+}n{+}4)$. Moreover, a  geometry, called the correlahedron, is defined and its canonical form is conjectured to give the bosonised supercorrelator.

The correlators exhibit a full permutation symmetry. This suggests that the correlahedron geometry be invariant under any permutation of the twistors $X_i$ up to the action of a sign flip operator $F_I$. In other terms, for each permutation $\sigma\in S_n$ there must be a flip transformation $F_\sigma$ such that $ F_{\sigma} \sigma$ leaves  the correlahedron invariant. The set of all $\tilde \sigma = F_{\sigma} \sigma$ defines a group we call the signed symmetric group or signed permutation group.  

Just as for the amplitude, the correlator $\mathcal{G}_{n,k}$ is composed of two types of bracket. The $n{+}k{+}4$ brackets involving only $X$s and the uplifted conformal invariants $\braket{YX_iX_j}$.
We are interested now in classifying all permutation invariant geometries that are defined using these two types of brackets. The main result of this analysis will be that, for $k=n-4$, there exists just one class of geometries defined using $\braket{YX_iX_j}$ which can be represented by the correlahedron.

The maximally nilpotent  case $k=n-4$ presents the advantage that there is a unique bracket involving $X$s only, $\braket{X_1\cdots X_n}$ and we can always fix it to be positive. Because of the permutation symmetry, we can then choose an arbitrary bracket, such as $\braket{YX_1X_2}$, and use the action of the signed symmetric group to generate all the other brackets. By flipping  $Y\to -Y$ if necessary, we can fix $\braket{YX_1X_2}>0$. From this moment on we  will  indicate $\braket{YX_iX_j}$ with $\braket{X_iX_j}$ to make the notation more compact unless there is possible ambiguity.

Since we have already studied cyclic invariance in detail to classify the inequivalent amplituhedron-like geometries, we already know we can always choose representatives invariant under cyclic or twisted cyclic symmetry.  Applying powers of the cyclic permutation on $\braket{X_1X_2}>0$ we obtain
\begin{align}
	&\braket{X_iX_{i+1}}>0, &\text{for cyclic}, \nonumber \\
	&\braket{X_iX_{i+1}}>0, \qquad \braket{X_nX_1}<0  &\text{for twisted cyclic}.
\end{align}
Let us now consider the action under a second permutation, the transposition $(1,2)$. This operator can come in general with a flipping sign operator $F_I$, but not all sign strings $s$ are allowed. The transposition $(1,2)$ leaves invariant all brackets that do not contain $X_1$ or $X_2$ and the bracket $\braket{X_1X_2}$ itself. Therefore $F_s$ must act trivially on these brackets. The solutions for $F_s$ are 
\begin{align}
	I=\{\}, \quad I=\{1,2\}.\quad I=\{1,\cdots,n\},\quad I=\{3,\cdots,n\}.
\end{align}
The last two solutions are in fact equivalent to the first two and therefore there can  only be two types of transposition, $(i,j,+)=(i,j)$ and $(i,j,-)=F_{\{i,j\}}(i,j)$.

We can prove that a signed cyclic geometry invariant under $(1,2,\pm)$ is also invariant under the whole signed symmetric group. In fact, because of cyclicity, it will also be invariant under $(i,i+1,\pm)$ and the set of adjacent transpositions generates the symmetric group. This can be proven using the relation
\begin{align}
	(i,j)(j,k)(i,j)=(i,k),
\end{align}
or more specifically 
\begin{align}
	(i,i+1)(i+1,i+l)(i,i+1)=(i,i+l+1).
	\label{genL}
\end{align}
Therefore if we start with $(i,i+1,+)$ and $l=0$ we can then use \eqref{genL} to generate all permutations. The resulting inequalities defining the geometry will read
\begin{align}
	\braket{X_iX_j}>0, \qquad \braket{X_1 \cdots X_n}>0.
	\label{symmetric}
\end{align}
In particular we obtain that $\braket{X_1X_n}>0$, therefore the geometry generated by $(1,2,+)$ can only be cyclic and not twisted cyclic. 

If on the other hand the geometry is invariant under $(i,i+1,-)$ instead, we can see that
\begin{align}
	(i,i+1,-)(i+1,i+2,-)(i,i+1,-)=(i,i+2,+),
\end{align}
from which we derive that
\begin{align}
	(i,i+1,-)(i+1,i+l,(-1)^{l-1})(i,i+1,-)=(i,i+l,(-1)^{l}).
	\label{comp}
\end{align}
Therefore if a geometry is invariant under $(i,i+1,-)$, for all $i$ except $i=n$, then \eqref{comp} tells us it must also be invariant under $(2,n,(-1)^n)$. If we act with $(2,n,(-1)^n)$ on $\braket{X_1X_2}>0$ we obtain 
\begin{align}
	(2,n,(-1)^n)\braket{X_1X_2}>0\quad \Rightarrow\quad(-1)^n\braket{X_1X_n}>0.
\end{align}
This implies that geometries generated by negative transpositions must be cyclic for $n$ odd and twisted cyclic for $n$ even. Therefore for fixed $n$ we just have two types of geometry:  one invariant under positive adjacent transpositions and one invariant under negative adjacent transpositions. The geometry invariant under $(1,2,+)$ is described by \eqref{symmetric}, while the one invariant under $\braket{1,2,-}$ is described by the following inequalities
\begin{align}
	&\braket{X_1\cdots X_n}>0, \nonumber \\
	&(-1)^{l+1}\braket{X_iX_{i+l}}>0,  &\text{for } i+l \leq n, \nonumber\\
	&(-1)^{n+l+1}\braket{X_iX_{i+l}}>0, &\text{for } n<i+l<2n.
\end{align}
At this point we can still use $F_\text{even}$ or equivalently $F_\text{odd}$ to see if these two set of inequalities are actually equivalent. Representing the action of $F_\text{even}$ on the brackets we obtain
\begin{align}
	&F_\text{even}\braket{X_iX_{i+l}}=\braket{X_iX_{i+l}}(-1)^{l+1}, \qquad &\text{for } i+l \leq n \nonumber \\
	&F_\text{even}\braket{X_iX_{i+l}}=\braket{X_iX_{i+l}}(-1)^{l+n+1}, \qquad &\text{for } n<i+l<2n 
\end{align}
The $F_\text{even}$ or $F_\text{odd}$ operator maps a set of inequalities invariant under $(1,2,+)$ to one invariant under $(1,2,-)$. Moreover it maps cyclic to twisted cyclic for $n$ odd. Therefore for any $n$ the geometry compatible with the bosonized maximally-nilpotent correlator is unique and can be described by \eqref{symmetric}.

\section{Conclusions and Outlooks}

In this work we have used the topological characterization of the amplituhedron in terms of flipping numbers to study the geometry of the squared amplituhedron. In this new language the amplituhedron is defined as the geometry having maximal flipping numbers and positive proper boundaries, up to the one fixed by twisted cyclicity. The squared amplituhedron corresponds instead to the union of all geometries without restriction on the flipping numbers and positive proper boundaries.
We named the geometries with non-maximal flipping number  amplituhedron-like geometries and propose that these correspond to products of amplitudes (in the case of minimal number of points $n$) giving  proofs of this at tree-level and MHV loop level.
We have given an alternative  non-intrinsic characterisation of the geometries (at tree and loop level)  and  their natural triangulation as sums of  pairs of on-shell diagrams (at tree level).

While the superamplitude has maximal residues equal to $\pm1$ the square of superamplitude  has maximal residues in  $ 2 {\mathbb{Z}}$.  We identify in the structure of the maximal squared amplituhedron a geometrical interpretation of this feature. In fact we have found that the amplituhedron-like geometries that compose the squared amplituhedron are almost disconnected, which means that their interior are disconnected but their boundaries intersect on regions of codimension smaller then $1$. Each almost disconnected component is a positive geometry and therefore has a canonical form with maximal residues equal to $\pm1$.  Maximal residues with value higher then $1$, correspond to points in the Grassmannian where these almost disconnected geometries touch.  A generalisation of canonical form, the  (globally oriented) canonical form can act on such unions of almost disconnected positive geometries, and when acting on the squared amplituhedron gives the square of the amplitude. This square will involve a sum over pairs of equivalent geometries that is responsible for the factor of $2$ in all maximal residues of the superamplitude squared.

The geometries corresponding to individual flipping ``patterns'' will in general have spurious boundaries which only cancel in the sum. However the case of flipping number $f=k/2$ may be an interesting exception to this rule. At six points,  $\sH_{6,2}^{(1)}$ is given by two sign flip patterns $\{+,+,-\}$ and $\{+,-,-\}$ which are equivalent under the $F_{\text{odd}}$ map and therefore each have the same canonical form, which is equal  to  $1/2(A_{6,1})^2$ \eqref{A61}. From the denominator of $(A_{6,1})^2$ we can immediately see that $\braket{1235}$ is not a boundary of the two flipping patterns, unlike for $A_{6,1}$ itself. The two sign flip patterns are disconnected and the globally oriented canonical form of the union of these gives  the expected result,  $2A_{6,1}$, while the standard canonical form of the union vanishes.
More generally for $f=k/2=(n-4)/2$ (at tree level) the equivalence relation \eqref{dualgeometry} maps the geometry into itself and maps flipping patterns pairwise into each other. 
This should be contrasted with  $\sH_{7,3}^{(1)}$ and  $\sH_{7,3}^{(2)}$, for which $f \neq k/2$. 
Here each flipping pattern has spurious poles $\braket{1235},\braket{1236}$ which cancel in the sum and the union gives a connected geometry.
It would be interesting to verify if this persists in general, that is  $\sH_{n,n-4}^{(f)}$ is a connected geometry for $f\neq \frac{k}{2}$, whereas for $f= \frac{k}{2}$ it has two disconnected components.

One would like to investigate  further generalizations of the amplituhedron geometry and what they correspond to.
The most obvious thing is to  consider geometries defined in a   similar way to the amplituhedron-like geometries but with different signs for the inequalities. This seems to immediately lead to a vast number of cases. However restricting to non-equivalent geometries and imposing cyclicity reduces the number of possibilities. As shown in section~\ref{CanonicalizingCyclicity} this reduces to examining cyclic (or twisted cyclic) geometries. So to be concrete we could imagine considering more general choices for the physical inequalities which at the moment we take to be $\braket{Yii{+}1jj{+}1}>0$. We could generalise by imposing different signs for the inequalities $\braket{Y12jj{+}1}\lessgtr 0$ (with the sign of $\braket{Yii{+}1jj{+}1}>0$ for other $i$ following by (twisted) cyclicity). One needs to first examine if it makes sense to have some the analogue of flipping patterns etc in these cases.
In any case it would be interesting to examine such geometries and understand what they correspond to. 

Objects we might imagine appearing from more general geometries of this type are  
products of more than two amplitudes, the simplest example would be NMHV${}^3$ at 7 points. These functions are well defined cyclic dlog forms, which can be seen by their expression as products of on-shell diagrams. As observed in \cite{Arkani-Hamed2018}, for $m=2$ the N${}^k$MHV amplituhedron is a product of NMHV amplituhedra, $\cA_{n,k}=\frac{(\cA_{n,1})^{k}}{k!}$ and as noted below~\eqref{129} we find by explicit computation that in fact all flipping {\em patterns} of maximal amplituhedron-like geometries $\sH_{n,n-2,2}^{(f)}$ have canonical form equal to $A_{n,k}$. For $m=4$ the generalization is still unclear.

If we go beyond the case of maximal $k=n-m$ then even more possible natural generalisations emerge. Going beyond this case, the natural correspondence between amplituhedron-like geometries and the product of amplitudes does not seem to hold in the same way.
For example we have checked by direct calculation that for  $k=1$, $n=6$ 
the direct generalization of \eqref{conjecture1}  is not true,
\begin{align}
	H_{n,k}^{(f)}\neq A_{n,f}*A_{n,1-f} \qquad \text{for }n=6,\ k=1\ .
\end{align} 
Therefore two immediate questions arise: 
\begin{enumerate}
	\item What does the amplituhedron-like geometry correspond to for $n>k+m$?  
	\item Is there a geometry corresponding to the product of two general (i.e. non parity conjugate) amplitudes?
\end{enumerate}
We have an answer to the first question in the MHV case at loop level as the product of MHV and $\overline{\text{MHV}}$ amplitudes~\eqref{sqrr} 
\begin{align}\label{sqrr2}
	H_{n,0,l}= \sum_{l'} \overline{A}_{n,0,l'}\, A_{n,0,l-l'}\ .
\end{align}
However, in the above equation $\overline{A}$ is the anti-MHV amplitude divided by tree-level. Such a quantity has no analogue beyond the  MHV case
and so it is not clear how this formula will generalise beyond this case.

In looking at the second question note that we have checked numerically that, for $k<n-m$,  the alternative characterisation of amplituhedron-like geometries~\eqref{GrassmannianProduct}
no longer works: it is no longer equivalent to the amplituhedron-like geometries. Nevertheless, the association of a geometry to the product of on-shell diagrams described in \eqref{pi} could be a starting point for a systematic derivation of  the geometry corresponding to the products   $A_{n,f}*A_{n,k-f}$ in a similar way as has been done in \cite{Herrmann:2020qlt} for chiral pentagon integrands. In that case, the requirement that the geometry of chiral pentagons giving the 1 loop MHV amplitude must share spurious co-dimension 1 boundaries isolates a unique solution of the geometry of the chiral pentagon.  It would be interesting to see if similar constraints identify a unique geometry for the product of amplitudes.

From the geometric point of view the non-trivial nature of the external $Z$ space in the non-maximal case opens up many new possibilities by looking at non-convex geometries. Instead of imagining the $Z$s as fixed and looking at the space of possible  $Y$s we could just as well consider varying the $Z$s. Convexity of the $Z$s  can be rewritten in a suggestive way using sign flip language~\cite{Arkani-Hamed2018}
\begin{align}\label{Zpositivity}
	&\braket{i_1 (i_1+1) \cdots i_{\frac{k+4}{2}}(i_{\frac{k+4}{2}}+1)}>0, \ \ \braket{i_1 (i_1+1) \cdots i_{\frac{k+2}{2}}(i_{\frac{k+2}{2}}+1) 1 n}>0 \quad  \text{for }k \text{ even}, \nonumber \\
	&\braket{1i_1 (i_1+1) \cdots i_{\frac{k+3}{2}}(i_{\frac{k+3}{2}}+1)}>0, \ \ 
	\braket{i_1 (i_1+1) \cdots i_{\frac{k+3}{2}}(i_{\frac{k+3}{2}}+1)  n}>0 \quad   \text{for }k \text{ odd}, \notag\\
	&\{\braket{123\cdots (k{+}3) i}\}  \qquad i=k{+}4,..,n \quad  \text{ has no sign flips}\,.
\end{align}
Even though this is clearly a small subset of all the possible ordered $Z$ determinants, these constraints alone are sufficient to imply that $Z\in Gr_{>}(k{+}4,n)$.
This suggests considering generalised geometries where the last line is replaced by a flipping number condition for the external $Z$ data, $f_Z$ where for the  amplituhedron, $f_Z=0$. 
These types of geometry could still in principle arise from the geometric light-like limit of the correlahedron. Recall the correlahedron is the space of $Y\in Gr(k{+}n,k{+}n{+}4)$ constrained by the equations $\braket{YX_iX_j}>0$.  The geometry of the squared amplituhedron  can be derived from the correlahedron by imposing the constraints $\braket{YX_iX_{i+1}}=0$ and then projecting respect to the intersection points $\tilde Y_i=Y\cap (X_iX_{i+1})$.   The external data, $Z$, emerges as the points  $Z_i =X_i\cap X_{i+1}$ on $\tilde Y^\perp$, while $Y$ can be rewritten as $Y=\tilde Y \hat Y$, where the allowed values of $\hat Y$ gives the squared amplituhedron. In \cite{Eden2017} the  squared amplituhedron geometry, defined as in~\eqref{sqamp} was derived from the geometric light-like limit of the correlahedron. However whereas in the maximal case the $Z$ space is unique,  in the non-maximal case this is no longer true and non convex $Z$s could arise from the light-like limit.

In conclusion, more work is needed to derive the geometry corresponding to products of amplitudes in general as well as the related non-maximal squared amplituhedron geometry. However we suspect non-convex $Z$ configurations could indeed appear.

One could also consider more general geometries still by initially relaxing the assumption of manifest cyclicity. 
Such  types of non cyclic geometries have been explored for $k=2$ and $m=2$ in \cite{Herrmann:2020qlt}. They introduced flipping patterns for the physical inequalities themselves and showed that the resulting geometries  correspond to interesting loop integrands called chiral octagons. 
Remarkably, by taking the union of many of these types of non-cyclic geometries, they obtain  
a new geometry, not equal to the amplituhedron,  but which nevertheless gives the  one loop MHV amplitude. In this construction therefore cyclicity emerges in a less trivial manner by taking the cyclic sum of non-cyclic geometries.

It would also be interesting to explore if a similar connection between non maximal flipping number and products of parity conjugate amplitudes holds in the context of the momentum amplituhedron~\cite{Damgaard:2019ztj}, where its definition in terms of the sign flip number can be naturally generalized.

Finally there is a huge amount still to be explored and understood with regard to correlators and the correlahedron and its interaction with amplitudes and the amplituhedron. Can the geometry be put to practical use in determining amplitudes or correlators at higher loops. An example of an idea in this direction is the deepest cut~\cite{Arkani-Hamed2019} which gives predictions for non-trivial residues of amplitudes  at high loop order.

\subsection*{Acknowledgements}

We thank Davide Polvara for comments on an earlier draft of this work. 
This project has received funding from the European Union’s Horizon 2020 research and innovation programme under the Marie Sklodowska-Curie grant agreement No.764850 ``SAGEX”. PH is supported by STFC Consolidated Grant ST/P000371/1.

\medskip

\appendix
\section{Star product proof for $m=1$ }

From~\eqref{prodRule2}  we need to prove that when we put $Y=Y_0$ and $Z=Z(\chi)$\eqref{zs} then
\begin{align}
	\frac{1}{N(k_1,1)}\braket{I}_{k_1}\frac{1}{N(k_2,1)}\braket{J}_{k_2}=(-1)^{k_1k_2+k_2}\frac{1}{N(k_1+k_2,1)}\braket{I(Y\cap J))}\,.
	\label{prodm1}
\end{align}
Explicitly we have $N(k,1)=(-1)^{\lfloor k/2 \rfloor}k!$,  $Y= \begin{pmatrix}0&0\\\mathbb{1}_{k_1\times k_1}&0\\0&\mathbb{1}_{k_2\times k_2} \end{pmatrix}$ and each bosonised momentum twistor in $k{+}m$ space is given as $Z_i=(z_i, \chi_i\phi_1,\cdots,\chi_i\phi_k)$. Defining $Y_1=(0,\mathbb{1}_{k_1\times k_1},0)$ and $Y_2=(0,0,\mathbb{1}_{k_2\times k_2})$, this becomes
\begin{align}\label{expprod1}
	\frac{1}{N(k_1,1)}\braket{I}_{k_1}\frac{1}{N(k_2,1)}\braket{J}_{k_2}=  \frac{(-1)^{\frac{\lfloor k_1 \rfloor}{2}+\frac{\lfloor k_2 \rfloor}{2}}}{k_1!k_2!}\braket{IY_2}((-1)^{k_1}\braket{Y_1J})\ .
\end{align}
To obtain  this expression  we can expanded  $(Y\cap J)$ on $Y$ as 
\begin{align}\label{YinterExp}
	Y\cap J=\frac{1}{k_1!k_2!} Y_{a_1}\cdots Y_{a_{k_2}}\braket{Y_{a_{k_2+1}}...Y_{a_{k_1+k_2}} J} \epsilon^{a_1,...,a_{k_1+k_2}},
\end{align}
to obtain
\begin{align}
	\braket{I (Y\cap J)}=\frac{1}{k_1!k_2!}\braket{I Y_{a_{k_2+1}}...Y_{a_{k_1+k_2}} }\braket{ Y_{a_1}\cdots Y_{a_{k_2}}J} \epsilon^{a_1,...,a_{k_1+k_2}},
\end{align}
which shows how for $m=1$ the star product corresponds to nothing more than writing the simplest $SL(k)$ invariant formula that combines     $Y$,  $I$ and $J$ and has the correct scaling.
The role of the $Y$s as columns of the identity matrix is just to select the rows, and therefore the $\phi$s, entering the determinant. But since the $\phi$s are dummy variables that can be relabeled if we antisymmetrise respect to $Y=Y_1Y_2$ in \eqref{expprod1} we leave the expression unchanged. We can therefore rewrite the latter as
\begin{align}
	\frac{(-1)^{\frac{\lfloor k_1 \rfloor}{2}+\frac{\lfloor k_2 \rfloor}{2}{-}k_1}}{k!}(-1)^{k_1k_2}\braket{I(Y\cap J)},
\end{align}  
where $(-1)^{k_1k_2}$ come from the convention we chose for the sign of the intersection in equation \eqref{intersection}. Now we can use that $\frac{\lfloor k_1 \rfloor}{2}+\frac{\lfloor k_2 \rfloor}{2}+\frac{\lfloor k_1{+}k_2 \rfloor}{2} =k_1+k_2 \text{ mod } 2 $ and conclude that
\begin{align}
	\frac{1}{N(k_1,1)}\braket{I}_{k_1}\frac{1}{N(k_2,1)}\braket{J}_{k_2}= \frac{(-1)^{k_1k_2{+}k_2}}{N(k,1)}\braket{I(Y\cap J)}
\end{align}
which proves the star product formula for $m=1$.

\section{Bosonised product checks for $m=2$ and $m=4$}

Here we would like to give evidence for the star product rule~\eqref{prodRule} by explicitly computing both sides of the equation for some special cases and verify that they match. We have chosen two examples that highlight how the sum over permutations in \eqref{prodRule} is necessary to give the right result.

Consider the the following product of bosonised brackets for $m=2$ 
\begin{align}
	\left( \braket{123}\braket{234}\right) *\left( \braket{123}\braket{234}\right) = -\frac{1}{2}\braket{Y23}^2\braket{1234}^2.
\end{align}
We can verify this result by using \eqref{prodRule2} and projecting both sides on a pair of on-shell Grassmannian variables $(\chi_i)^2(\chi_j)^2$. If we project on $(\chi_1)^2(\chi_4)^2$ for example, we obtain for the left hand side
\begin{align}
	\frac{1}{N(1,2)^2}\int d^2 \phi_1d^2 \phi_2\left(\braket{23}^2 \phi_1\chi_1\phi_1\chi_4\right)\left(\braket{23}^2 \phi_2\chi_1\phi_2\chi_4\right)= \nonumber \\
	=\braket{23}^4\left(\frac{\braket{\chi_1\chi_4}}{2!}\right)^2=-\frac{1}{8}\braket{23}^4\braket{\chi_1\chi_1}\braket{\chi_4\chi_4}.
\end{align}
While for the right hand side we obtain
\begin{align}
	-\frac{1}{2}\braket{23}^4\frac{1}{N(2,2)}\int d^2\phi_1 d^2\phi_2 (\phi_1\chi_1\phi_2\chi_2-\phi_1\chi_2\phi_2\chi_1)^2=-\frac{1}{8}\braket{23}^4\braket{\chi_1\chi_1}\braket{\chi_4\chi_4}.
\end{align}
The two projections match as expected.

Let's  now see an example of product of bosonized brackets for $m=4$.
Consider the following product of bosonized brackets 
\begin{align}\label{prodex}
	\left( \braket{12367}^3\braket{12357}\right)*\left(\braket{134567}\braket{124567}^3\right)=\notag\\
	=
	\frac{1}{4}\left(\braket{Y1267}^2(\braket{Y1 2 6 7} \braket{Y1 3 5 7} + 
	3 \braket{Y1 2 5 7} \braket{Y1 3 6 7}\right)\braket{1234567}^4.
\end{align}
To check this relation we can again use \eqref{prodRule2} and project \eqref{prodex} on $(\chi_3)^4(\chi_4)^4(\chi_5)^4$. Projecting on $(\chi_i)^4$ is equivalent to acting with the operator $\partial^{(4)}_i:=\partial^{(4)}_{\chi_i}$. Projecting on the right and side and integrating out the $\phi$s its easy and gives 
\begin{align}\label{rightprojm4}
	\frac{1}{4} \braket{1267}^6 \left ( \braket{Y1 2 6 7} \braket{Y1 3 5 7} + 
	3 \braket{Y1 2 5 7} \braket{Y1 3 6 7}\right)
\end{align}If we perform the same operation on the left hand side instead we obtain
\begin{align}
	\partial^{(4)}_3\partial^{(4)}_4\partial^{(4)}_5([45]^3[46][2][3]^3)=\nonumber \\ 
	= \partial^{(3)}_3[45]^3 \partial_4^{(3)}\partial_5^{(3)}[3]^3\left( \partial_3^{(1)}[46] \partial_4^{(1)}\partial^{(1)}_5[2]+\partial_5^{(1)}[46]\partial^{(1)}_3\partial_4^{(1)}[2]\right)= \nonumber \\
	\braket{3333}_\phi[345]^6\left([346][245]\braket{1111}_\phi\braket{2222}_\phi+[456][234]\braket{1222}_\phi\braket{1112}_\phi\right)+ (\phi_2\leftrightarrow \phi_3),
\end{align}
where $\braket{ijlk}_\phi= \epsilon_{ABCD} \phi_i^A\phi_j^B\phi_l^C\phi_k^D$ and $[ij]$ indicate a 5-bracket not containing indices $i,j$ and analogously $[ijk]$ indicates a 4-brackets not containing the indices $i,j,k$ . Manipulating the expression using the identities
\begin{align}
	\braket{a***}_{\phi}\braket{aaa*}_{\phi}= -{\binom{4}{3}}^{-1} \braket{aaaa}_{\phi}\braket{****}_{\phi}\\
	\braket{aa**}_{\phi}\braket{aa**}_{\phi}= {\binom{4}{2}}^{-1}\braket{aaaa}_{\phi}\braket{****}_{\phi}
\end{align}
and integrating out the $\phi$'s we obtain
\begin{align}
	[345]^6\left([346][245]-\frac{1}{4}[456][234]\right)=\braket{1267}^6 \left( \braket{1257}\braket{1367}-\frac{1}{4}\braket{1237}\braket{1567}\right),
\end{align}
which can be tested numerically to be equal to \eqref{rightprojm4} as expected.

 \providecommand{\href}[2]{#2}\begingroup\raggedright
\end{document}